\newcommand{\diracslash}[1]{#1\llap{/\kern2pt}}

\newcommand{\be}{\begin{equation}}
\newcommand{\ee}{\end{equation}}
\newcommand{\bea}{\begin{eqnarray}}
\newcommand{\eea}{\end{eqnarray}}
\newcommand{\ba}[1]{\begin{array}{#1}}
\newcommand{\ea}{\end{array}}

\newcommand{\bt}{\begin{tabular}}
\newcommand{\et}{\end{tabular}}

\newcommand{\beas}{\begin{eqnarray*}}
\newcommand{\eeas}{\end{eqnarray*}}

\documentclass[preprint,prd,aps,floats,nofootinbib,floatfix]{revtex4}

\DeclareSymbolFont{rsfs}{U}{rsfs}{m}{n}
\DeclareSymbolFontAlphabet{\mathrsfs}{rsfs}
\usepackage{graphicx}
\usepackage{multirow}

\usepackage{aliascnt}

\usepackage{pgf}

\usepackage{pdfpages}
\usepackage{amsmath}
\usepackage[amsmath,thmmarks]{ntheorem}

\usepackage{graphicx}
\usepackage{hyperref}
\usepackage{cleveref}

\begin{document}


\title{In-medium pseudoscalar  $D/B$  mesons and charmonium decay width}

\author{Rahul Chhabra}
\email{rahulchhabra@ymail.com,}
\affiliation{Department of Physics, Dr. B R Ambedkar National Institute of Technology Jalandhar,
 Jalandhar -- 144011,Punjab, India}

\author{Arvind Kumar}
\email{iitd.arvind@gmail.com, kumara@nitj.ac.in}
\affiliation{Department of Physics, Dr. B R Ambedkar National Institute of Technology Jalandhar,
 Jalandhar -- 144011,Punjab, India}

\def\be{\begin{equation}}
\def\ee{\end{equation}}
\def\bearr{\begin{eqnarray}}
\def\eearr{\end{eqnarray}}
\def\zbf#1{{\bf {#1}}}
\def\bfm#1{\mbox{\boldmath $#1$}}
\def\hf{\frac{1}{2}}
\def\kp{\zbf k+\frac{\zbf q}{2}}
\def\km{-\zbf k+\frac{\zbf q}{2}}
\def\hwo{\hat\omega_1}
\def\hwt{\hat\omega_2}

\begin{abstract}

Using QCD sum rules and chiral SU(3) model, we investigate the effect of temperature, density, strangeness fraction and isospin asymmetric parameter on the shift in masses and decay constants of the pseudoscalar $D$ and $B$ meson in  hadronic medium, which consist of nucleons and hyperons. The in-medium properties of $D$ and $B$ mesons within QCD sum rule approach depend upon the quark and gluon condensates. In chiral SU(3) model, quark and gluon condensates are introduced through the explicit symmetry breaking term and the trace anomaly property of the QCD, respectively and are written in terms of scalar fields $\sigma$, $\zeta$, $\delta$ and $\chi$. Hence, through medium modification of $\sigma$, $\zeta$, $\delta$ and $\chi$ fields, we obtain the medium modified masses and decay constants of $D$ and $B$ mesons. As an application, using $^3 P_0$ model, we calculate the in-medium decay width of the higher charmonium states  $ \psi(3686)$, $\psi(3770)$ and $\chi(3556)$ to the $D \bar{D}$ pairs, considering the in-medium mass of $D$ mesons. These results may be important to understand the possible outcomes of high energy physics experiments, e.g., CBM and PANDA at GSI, Germany.

\textbf{Keywords:} Dense hadronic matter, strangeness fraction,
 heavy-ion collisions, effective chiral model, QCD sum rules,  heavy mesons.

PACS numbers : -14.40.Lb ,-14.40.Nd,13.75.Lb
\end{abstract}

\maketitle
\section{Introduction}
\label{sec_intro}
Quark gluon plasma, which is believed to be the free state of quarks and gluons is one of the most interesting topic of present day hadronic physics. Full understanding of this state can reveal the hidden myth behind the origin of our present day universe and transformation of QGP phase to the hadronic phase. On-going heavy ion collision experiments e.g, Large Hadron Collider (LHC),  Relativistic Heavy Ion Collider (RHIC) and the future  Compressed Baryonic Matter (CBM) experiment are intended to explore the QCD phase diagram at different values of temperature and density. The LHC and RHIC experiments which work on the condition of high temperature and low dense medium and is compliment to future experiment, CBM, have interpreted the production of QGP \cite{exp,meth}. At such extreme conditions, hadrons melt and quarks and gluons can roam freely in the medium. Direct investigation of this state is difficult due to its existence for short interval of time \cite{ramona}.

 Many proposals have been developed to observe the creation of QGP in heavy ion collision experiments \cite{ramona}. Famous signals of QGP are strangeness enhancements in heavy ion collisions as suggested by the  Capella and Soff \cite{capella, soff}, jet quenching in the parton energy loss formulated by the Bjorken \cite{bjorken}, enhancements in dileptons spectra in nucleus-nucleus collisions  observed in HELIOS and DLS \cite{helios, DLS}, and in heavy ion collisions \cite{srivastava}. Beauty of the dileptons is the least interaction with the medium and these dileptons can give us  an undistorted information about QGP medium.
  Apart from these, the phenomenon $J/ \psi$ suppression observed
in heavy-ion collision experiments \cite{NA,rhic1,rhic2,alice1,alice2} can also be a signature of the formation of QGP as was proposed  first by Matsui and Satz \cite{satz}.
   However, not all of the observed  $J/ \psi$ suppression in nucleus-nucleus  collisions  is due to QGP formation \cite{bramb}. As mentioned in \cite{lei}, $J/ \psi$ suppression may be because of co-movers interactions or by inelastic scattering of $J/ \psi$ with surrounding nucleons \cite{gers}. Nuclear dependence of $D$ mesons can also alter the $J/ \psi$ suppression observed by  NA50 collaboration \cite{NA, gore, zhang, cassing1, cassing2}. In a similar way, nuclear dependence of $B$ mesons can also interfere the $\Upsilon$ suppression in heavy ion collision experiments. 
 
 In addition to $J/ \psi$ suppression, in-medium properties of $D$ mesons can also reveal about the existence of $D-$mesic nuclei \cite{tsu2}. Another peculiar thing about the $D$ meson is about its mass which is 1869 MeV in vacuum  and it is much more than the sum of the masses of its constituents which depicts that the Higgs mechanism of mass generation in the standard model is of little importance in an explanation of the mass of the matter around us \cite{hilgert}.
 The excess mass of hadrons are considered as a result of interaction of quarks and gluons with the ground state of QCD  which is populated with quark and gluon condensates, and to spontaneous breaking of chiral symmetry. 
 To understand the medium modification of  $D$ mesons it is important to understand their production and its collective flow in the medium. It became possible due to the experimental facility at Jefferson Lab, USA \cite{jlab}, in which CEBAF accelerator is used to produce the continous beam of electrons and these are scattered off to produce charm hadrons. At the upcoming FAIR (Facility for Antiproton and Ion Research) project, in PANDA experiment, with annihilation of antiprotons on nuclei, whereas in CBM, with the use of Au nuclei, intentions are to explore the properties of open charm mesons.

Theoretically, many methodologies have been developed to study the in-medium properties of $D$ and $B$ mesons. In \cite{tsu2} authors  used quark meson coupling model (QMC) and observed a negative mass shift of the $D$ meson in the nuclear medium. In QMC model, $D$ meson is considered as bound state of one light quark and one charm quark. The interaction of $D$ meson  with the nucleons occur through the exchange of the scalar and vector mesons.   In \cite{tolo2} using coupled channel $G$ matrix approach authors observed increase in the mass of $D$ meson in nuclear medium.
 The chiral hadronic SU(3) model
 is generalized to SU(4) and SU(5) sector,
 for investigating the in-medium properties of pseudoscalar
 $D$ and $B$ mesons \cite{arvind3,amdp1,amdp2}. The QCD sum rules were used in \cite{haya,hilger,azizi,wang2,scalarD,qcd2} to 
 evaluate the masses and decay constants of $D$ and $B$ mesons at finite density of nuclear matter. Within QCD sum rules, the in-medium properties of heavy mesons are expressed in terms of
 quark and gluon condensates and  in-medium modification of these condensates lead to medium modification of the properties of heavy meson.
In \cite{arv1,rahul}, using QCD sum rules and chiral SU(3) model \cite{papa}, we investigated the properties of scalar, vector and axial-vector $D$ and $B$ mesons at finite density and temperature of the hadronic matter. The medium modified values of quark and gluon condensates were calculated using chiral effective SU(3) model and these were further used as input in the QCD sum rules to evaluate the properties of $D$ and  $B$ mesons. This strategy help us to calculate the properties of heavy mesons at finite density and temperature of medium, for different values of isospin asymmetry and strangeness fraction.
 In the present paper, our objective is to calculate the in-medium masses and decay constants of pseudoscalar $D$ and $B$ mesons by using again chiral SU(3) model and QCD sum rules \cite{wang2}. Further, we will use the in-medium masses of pseudoscalar $D$ mesons to calculate the in-medium partial decay width of higher charmonium states $\chi(3556)$, $\psi (3686)$, $\psi (3770)$  to $D \bar{D}$ pairs by using $^3P_0$ model \cite{friman} and shall discuss its possible implications on the $J/\psi$ suppression.

 Higher charmonium and bottomonium states are considered as main source of the $J/ \psi$  and $\Upsilon$ states, respectively. If the mass of $D$ ($B$) mesons decrease in the medium then higher charmonium and bottomonium states may decay to $D\bar{D}$ and $B\bar{B}$ pairs, respectively, instead of decaying to ground state charmonium (bottomonium) states. Therefore, the drop of mass of $D$ and $B$ mesons may cause $J/\psi$ and $\Upsilon$ suppression in heavy ion collision experiments. Thus, the decay of higher charmonium and bottomonium states play an important role for the better understanding of the non-perturbative regime of QCD. Many models have been developed in the past to study the decay widths of hadrons e.g., hadrodynamic model in which hadrons are described as elementary point-like objects \cite{R1}, elementary emission models in which mesons are treated as extended objects \cite{R2,R3} but the decays occur via elementary meson emission, and elementary meson emission model in which emitted meson is considered as an elementary particle coupled to the quark \cite{ferre, bonnaz}. 
 
 The study of the decay widths of higher charmonium states is also important as a fact that it can be directly measured through the spectrum of dileptons of $p \bar{A}$ and $A A$ reactions in different heavy ion collision experiments \cite{friman}, and this can be taken as a helpful tool to validate the phenomenological methods. To achieve this task many pair creation models were proposed like $^3S_1 $ model (quark-antiquark pair is created from the gluon emitted by a quark of the original meson) \cite{furui},  flux tube model \cite{kokoski} and $^3 P_0$ model in which  hadron decay proceed through the $q \bar{q}$ pair with vacuum quantum numbers i.e., $J^{PC}$ =  $0^{++}$ \cite{micu}. As far as production of quark and anti-quark is concerned in Ref. \cite{ei}
  time-like part of the vector Lorentz confining interaction was considered as main reason  for the $q \bar{q}$ production whereas, in Ref. \cite{ackleh,deng} the $q \bar{q}$ production was suggested through the gluon exchange and scalar confining interactions. Further,  the spin orbit splitting observed in the heavy quarkonium, predicted a scalar confinement potential \cite{dobb, lees}, while in the study of decay widths of $P-$wave $D$ mesons, the mixture of scalar and vector potential was used \cite{adachi,vij}. 

The outline of the paper is as follows: In \cref{sec_chiral_model}, we discuss the chiral SU(3) model to calculate the in-medium scalar fields and use it to compute in-medium quark and gluon condensates.
In \cref{sec_qcdsumrules}, we describe the QCD sum rules to solve the in-medium masses and decay constants of $D$ and $B$ mesons. In \cref{sec_3p0_model}, we briefly discuss $^3P_0$ model to calculate the medium effects on the decay of higher charmonium states $\psi(3686)$, $\psi(3770)$ and $\chi(3556)$ to the $D \bar{D}$ pairs, by considering the medium effects of the $D$ meson mass. In \cref{sec_results_discussions}, we present the results and discussions of the present work and finally in \cref{sec_summary} we shall give a brief summary.

 \section{Chiral SU(3) model}
 \label{sec_chiral_model}
Chiral SU(3) model is an effective theory applied in non-perturbative  regime of QCD \cite{papa}. It is based on the nonlinear realization and broken scale invariance properties of chiral symmetry  \cite{w}. Within model we have a general Lagrangian density which consists of a kinetic energy term, baryon meson interaction term which produce baryon mass, self-interaction of vector mesons which generates the dynamical mass of vector mesons, scalar mesons interactions which induce the spontaneous  breaking of chiral symmetry, and the explicit breaking term of chiral symmetry. This Lagrangian density can be solved by using mean field approximation under which only the scalar and vector fields contribute to the baryon meson interactions and for all other mesons the expectation values become zero \cite{papa,su3}. From the Lagrangian density, using Euler Lagrange equation of motion, we obtain coupled equations of motion for the scalar fields $\sigma$, $\zeta$, $\delta$  and scalar dilaton field $\chi$ \cite{rahul}. We solve these coupled equations of motion using mean field approximation for the different value of strangness fractions $f_s$, isospin asymmetric parameter $I$, temperature $T$, and, density $\rho_B$ of the medium. 

The strangeness fraction is defined as $f_s$ = $\frac{\Sigma_i |s_i|\rho_i}{\rho_B}$, here $s_i$ is the number of strange quarks, $\rho_i$ is number density of $i^{th}$ baryon and isospin asymmetric parameter is defined as $I$ = $\frac{\rho_n - \rho_p}{2\rho_B}$, here $\rho_n$ and $\rho_p$ denote the number density of neutrons and protons, respectively, and, $\rho_B$ is the total baryonic density \cite{arvind2}. To calculate the in-medium masses and decay constants of pseudo-scalar $D$ and $B$ mesons using QCD sum rules, we shall need to calculate the light quark condensates $\langle\bar{u}u\rangle$ and $\langle\bar{d}d\rangle$, and scalar gluon condensates $\langle \frac{\alpha_s}{\pi}G_{\mu\nu}^{a} G^{\mu\nu a}\rangle$. Using chiral effective model, we can express the scalar quark condensates in terms of scalar fields through the explicit symmetry breaking term. We have following expressions for the scalar condensates:
 \begin{align}
\left\langle \bar{u}u\right\rangle
= \frac{1}{m_{u}}\left( \frac {\chi}{\chi_{0}}\right)^{2}
\left[ \frac{1}{2} m_{\pi}^{2}
f_{\pi} \left( \sigma + \delta \right) \right],
\label{qu}
\end{align}
and
\begin{align}
\left\langle \bar{d}d\right\rangle
= \frac{1}{m_{d}}\left( \frac {\chi}{\chi_{0}}\right)^{2}
\left[ \frac{1}{2} m_{\pi}^{2}
f_{\pi} \left( \sigma - \delta \right) \right],
\label{qd}
\end{align}
where $m_u$ and $m_d$ are the masses of $u$ and $d$ quarks, having values as $5$ and $7$ MeV, respectively. 
Further, from the broken scale invariance property of QCD, we know that the trace of energy momentum
tensor is nonzero and is equal to the scalar gluon condensates  $\langle \frac{\alpha_s}{\pi}G_{\mu\nu}^{a} G^{\mu\nu a}\rangle$. We can caricature the trace anomaly in the effective chiral model through the scale breaking term of the effective Lagrangian density which can we further used to
evaluate the trace of energy momentum tensor. Comparing the trace of energy momentum tensor evaluated within chiral model with that of QCD, we can express the gluon condensates, $\langle \frac{\alpha_s}{\pi}G_{\mu\nu}^{a} G^{\mu\nu a}\rangle$ in terms of the scalar fields $\sigma$, $\zeta$, $\delta$, and $\chi$ as
\begin{align}
\left\langle  \frac{\alpha_{s}}{\pi} {G^a}_{\mu\nu} {G^a}^{\mu\nu}
\right\rangle =  \frac{8}{9} \Bigg [(1 - d) \chi^{4}
+\left( \frac {\chi}{\chi_{0}}\right)^{2}
\left( m_{\pi}^{2} f_{\pi} \sigma
+ \big( \sqrt {2} m_{k}^{2}f_{k} - \frac {1}{\sqrt {2}}
m_{\pi}^{2} f_{\pi} \big) \zeta \right) \Bigg ],
\label{glu}
\end{align}
where ($m_\pi$, $f_\pi$) and ($m_K$, $f_K$) represents the (mass, decay constant) of $\pi$ and $K$ meson, respectively.
The dilaton field $\chi$ is introduced to break the scalar invariance property of QCD. In \cref{glu} $\chi_0$ denotes its vacuum value, parameter $d$ is a constant having value $2/11$, determined through the QCD beta function at one loop level for three colors $N_c$ and three flavors $N_f$ \cite{papa}.      

 \section{QCD Sum rules for D and B mesons}
 \label{sec_qcdsumrules}
The present section is devoted to give an understanding of QCD sum rules used to investigate the masses and decay constants of pseudoscalar $D$ and $B$ mesons in isospin asymmetric strange hadronic matter at finite temperature and density \cite{wang2}. As we will see below the mass shift and decay shift of heavy pseudoscalar mesons will be expressed in terms of quark and gluon condensates.  
 In QCD sum rule, we start with two point correlation function $\Pi(q)$, which is
 the Fourier transformation of the expectation value of the time-ordered product of isospin averaged current $J_5$, i.e., we write \cite{wang2}
\begin{align}
\Pi (q) = i\int d^{4}x\ e^{iq_{\mu} x^{\mu}} \langle T\left\{J_5(x)J_5^{\dag}(0)\right\} \rangle_{\rho_B, T} 
\label{tw}
\end{align}
 where $ x^\mu = (x^0,\textbf{x})$ is the four coordinate and $q^\mu = (q^0,\textbf{q})$ is four momentum. The spin-isospin averaged current $J_5$ is defined by
\begin{align}
 J_5(x) &= J_5^\dag(x) =\frac{\bar{c}(x)i\gamma_5 q(x)+\bar{q}(x)i\gamma_5 c(x)}{2},
 \label{psc}
 \end{align}
where $q$ is for light quark $u$ or $d$ and $c$ is for charm quark. Note that we are averaging the current of particles and antiparticles i.e., $D$ and $\bar{D}$ mesons. This can be understood as follows: The $D$ meson isospin doublet consist of  $D^{+}$ and $D^{0}$ mesons whereas in 
$\bar{D}$ meson doublet we have $D^{-}$ and $\bar{D^{0}}$ mesons. 
The mesons $D^{+}$, $D^{-}$, $D^{0}$
and $\bar{D^{0}}$ have the quark compositions,
$c\bar{d}$, $d\bar{c}$, $c\bar{u}$ and $u\bar{c}$,
respectively. We see that $D^{+}$ and $D^{-}$ are particle-antiparticles and similarly, $D^{0}$
and $\bar{D^{0}}$ mesons. Thus, when we use $q = d$  in \cref{psc}, we will find average current of $D^{+}$ and  $D^{-}$ mesons and when $q = u$, this will be for $D^{0}$
and $\bar{D^{0}}$. Thus, in the present work we shall find the average mass shift and decay shift of $D$ and $\bar{D}$ mesons under centroid approximation \cite{haya,wang2,scalarD}. As we will see later, this will enable us to find the mass-splitting between the isospin doublets due to isospin asymmetry of the medium at finite density and temperature. 
The even and odd part of QCD sum rules were used in Ref. \cite{hilger} to calculate the mass splitting between $D$ and $\bar{D}$ mesons.
For $B$ meson, charm quark $c$ will be replaced by the bottom $b$ quark.

The two point correlation function given by \cref{tw} can be divided into the vacuum part, a static one-nucleon part and pion bath contribution, i.e.,
\cite{wang2,zsc,kwon}.
 \begin{eqnarray}
\Pi (q) &=&\Pi_{0} (q)+ \frac{\rho_B}{2m_N}T_{N} (q) + \Pi_{P.B.}(q)\,,
\label{pibn}
 \end{eqnarray}
 where $T_N (q)$ is the forward scattering amplitude \cite{wang2}.
The purpose of third term i.e., pion bath term was to consider the contribution of the finite temperature of the medium \cite{zsc,kwon}. In our present investigation we take the contribution of finite temperature of the medium through the temperature dependence of the scalar fields $\sigma$, $\zeta$, $\delta$ and dilaton field $\chi$ calculated through the chiral SU(3) model.
  We have successfully used above idea to calculate the temperature dependence  of the masses and decay constants of vector and axial-vector $D$ and $B$ mesons \cite{rahul}. The temperature dependence of the optical potentials of kaons, $D$ and $B$ mesons have been calculated
  using the scalar fields $\sigma$, $\zeta$, $\delta$ and dilaton field $\chi$ in \cite{isoamss,arvind2}.
  
 In the limit ${\bf q} \rightarrow 0$, the forward scattering amplitude $T_N(\omega,{\bf q})$  can be  related to forward $D-N$ scattering $T$-matrix\cite{koi} 
  \begin{equation}
 {{\cal T}_{D N}}(m_{D},0)= 8\pi(m_N + m_{D})a_{D}.
\end{equation}
 where $a_{D}$ is the $DN$ scattering length.
 The phenomenological spectral density $\rho(\omega,0)$ can be parametrised  into three unknown parameters $a$, $b$ and $c$ as given below \cite{koi},
\begin{align}
\rho(\omega,0) &= -\frac{1}{\pi}
 \mbox{Im} \left[\frac{{{\cal T}_{D N}}(\omega,{ 0})}{(\omega^{2}-
m_{D}^2+i\varepsilon)^{2}} \right]  \frac{f_D ^2 m_D ^4}{m_c ^2} + \cdots, 
\nonumber\\
 &= a\,\frac{d}{d\omega^2}\delta(\omega^{2}-m_{D}^2)
 +
b\,\delta(\omega^{2} - m_{D}^2) + c\,\theta(\omega^{2}-s_{0})\,.
\label{a1}
\end{align}
The first term in above equation denotes the double-pole term and corresponds to 
the on-shell effects of  T-matrices,

\begin{equation}
a=-8\pi(m_N + m_{D})
 a_{D}f_{D}^2 m_{D}^2\,.
\label{a2}
\end{equation}
The second term in \cref{a1} denotes  the single-pole term,
and corresponds to the off-shell (i.e. $\omega^2\neq m_{D}^2$) effects of $T$-matrices. The
third term denotes the continuum term or the  remaining effects,
where, $s_{0}$, is the continuum threshold and this define the scale below which
the continuum contribution vanishes \cite{kwon}.
The symbols $m_c$, $m_D$ and $f_D$  denote the mass of charm quark, mass of $D$ meson and decay constant of $D$ meson, respectively. 

The shift in mass and decay constant of  $D$ and $B$ mesons can be written as \cite{wang2}
 \begin{equation}
\delta m_D = 2\pi \frac{m_N + m_D}{m_N m_D} \rho_N a_D,
\label{masshift}
\end{equation}
  and 
 \begin{equation}
 \delta f_D =  \frac{m_c ^2}{2f_D m^4}\left(\frac{b \rho_N}{2m_N} - \frac{4 f_D^2 m_D^3 \delta m_D}{m_c ^2}\right),
 \label{decayshift}
\end{equation}
respectively.

From \cref{a2,masshift,decayshift} we see that to find the mass shift and decay shift, we need to find the unknown parameters $a$ and $b$.
The unknown parameters are eliminated by equating the Borel transformed forward scattering amplitude $T_N(\omega,0)$ on the phenomenological side with the Borel transformed forward scattering amplitude $T_N(\omega,0)$ in operator product expansion side. Finally, we  obtain a  relation between parameters $a$ and $b$ and the quark and gluon condensates given by \cite{wang2}
\begin{align}
& a \left\{\frac{1}{M^2}\exp\left(-\frac{m_{D}^2}{M^2}\right) - \frac{s_0}{m_{D}^4} \exp\left(-\frac{s_0}{M^2}\right)\right\}
+b \left\{\exp\left(-\frac{m_{D}^2}{M^2}\right) - \frac{s_0}{m_{D}^2} \exp\left(-\frac{s_0}{M^2}\right)\right\}\nonumber\\
-&  \frac{2m_N(m_H+m_N)}{(m_H+m_N)^2-m_{D}^2}\left(\frac{f_{D}m_{D}^2g_{DNH}}{m_c}\right)^2\left\{ \left[\frac{1}{M^2}-\frac{1}{m_{D}^2-(m_H+m_N)^2}\right] \exp\left(-\frac{m_{D}^2}{M^2}\right)\right.\nonumber\\
&\left.+\frac{1}{(m_H+m_N)^2-m_{D}^2}\exp\left(-\frac{(m_H+m_N)^2}{M^2}\right)\right\}\nonumber\\
=&-\frac{m_c\langle\bar{q}q\rangle_N}{2}\left\{1+\frac{\alpha_s}{\pi} \left[ 6-\frac{4m_c^2}{3M^2} \right.\right.
\left.\left.-\frac{2}{3}\left( 1-\frac{m_c^2}{M^2}\right)\log\frac{m_c^2}{\mu^2}-2\Gamma\left(0,\frac{m_c^2}{M^2}\right)\exp\left( \frac{m_c^2}{M^2}\right) \right]\right\} \nonumber\\
&\exp\left(- \frac{m_c^2}{M^2}\right) \nonumber\\
&+\frac{1}{2}\left\{-2\left(1-\frac{m_c^2}{M^2}\right)\langle q^\dag i D_0q\rangle_N +\frac{4m_c
}{M^2}\left(1-\frac{m_c^2}{2M^2}\right)\langle \bar{q} i D_0 i D_0q\rangle_N+\frac{1}{12}\langle\frac{\alpha_sGG}{\pi}\rangle_N\right\} \nonumber\\
&\exp\left(- \frac{m_c^2}{M^2}\right)\, .
\label{qcdsumdst}
\end{align}
%
The nucleon expectation values of quark and gluon condensates appearing in OPE side of above equation can be calculated by using
\begin{eqnarray}
\mathcal{O}_{\rho_{B}} &=&\mathcal{O}_{vacuum} +
4\int\frac{d^{3}p}{(2\pi)^{3} 2 E_{p}}n_{F}\left\langle N(p)\vert \mathcal{O}\vert N(p) \right\rangle+
3\int\frac{d^{3}k}{(2\pi)^{3} 2 E_{k}}n_{B}\left\langle \pi(k)\vert \mathcal{O}\vert\pi(k) \right\rangle \nonumber\\
& =& \mathcal{O}_{vacuum} + \frac{\rho_B}{2 m_N}\mathcal{O}_{N} + 
\mathcal{O}_{P.B.}, 
\label{operator1}
\end{eqnarray}
where  $\mathcal{O}_{\rho_{B}}$  gives the expectation value of the operator at finite baryonic density. 
The term $\mathcal{O}_{vacuum}$ stands for the vacuum expectation value of the
operator, $\mathcal{O}_{N}$ give us the nucleon expectation value
of the operator and $\mathcal{O}_{P.B.}$ denotes
the contribution from the pion bath at finite temperature.
Also, $n_B$ and $n_F$ are the thermal Boson and Fermion distribution functions and are given by$\left[ e^{E_{k}/T} - 1 \right]^{-1}$ and 
$\left[ e^{\left( E_{p}-\mu_{N}\right) /T} - 1 \right]^{-1}$,
respectively.
As discussed earlier, the finite temperature effects in the present investigation will be evaluated through the scalar fields and therefore
contribution of third term will not be considered \cite{arv1}. 
Thus, within chiral SU(3) model, we can find the values of
 $\mathcal{O}_{\rho_{B}}$ at finite density of the nuclear medium and hence can find
 $\mathcal{O}_{N}$ using
\begin{equation}
\mathcal{O}_{N} = \left[ \mathcal{O}_{\rho_{B}}  - \mathcal{O}_{vacuum}\right] \frac{2m_N}{\rho_B}.
\label{condexp}
\end{equation}
The values of light quark condensate $<q \bar{q}>_{\rho_{B}}$ and the gluon condensate $\left\langle  \frac{\alpha_{s}}{\pi} {G^a}_{\mu\nu} {G^a}^{\mu\nu}\right\rangle_{\rho_{B}}$  are calculated using \cref{qu,qd,glu}. Also, $<\bar{q g_s \sigma G q}>_{\rho_{B}}$ and $<\bar{q} i D_0 i D_0 q>_{\rho_{B}}$ can be approximated in terms of $<q \bar{q}>_{\rho_{B}}$ \cite{arv1,rahul}.
We differentiate \cref{qcdsumdst}  w.r.t $\frac{1}{M^2}$ to make one more equation, so that we can solve two coupled equations to eliminate the unknown parameters $a$ and $b$.
%
%
%
%
%
%
Shift in mass and decay constant of $B$ meson can also be calculated through same equations by simply replacing the mass of $D$ meson (charm quark) with mass of $B$ meson (bottom quark).


\section{The $^3 P_0$ model}
\label{sec_3p0_model}


Now we outline the $^3P_0$ model used in the present work to calculate the in-medium strong decay with of higher charmonium states into $D \bar{D}$ pairs \cite{micu,yo,yaouanc}.
The $^3P_0$ model is a quark-antiquark pair creation model in which various transitions
can be studied using the pair creation strength parameter $\gamma$ and oscillator parameter $\beta$
which are fitted to experimental values. As was said earlier, in the $^3P_0$ model $q\bar{q}$ pair is created in the vacuum which combine with the $q\bar{q}$ of parent meson $A$ at rest decaying to $B$ and $C$ mesons.
The invariant matrix element for the particular type of decay  $A$ $\rightarrow$ $B$ + $C$ is expressed as \cite{bonnaz,friman}

\begin{eqnarray}
M_{A \rightarrow BC}&  \propto &
\int d^3k_q  \phi_A(2k_q-2k_B) \phi_B(2k_q-k_B) \phi_C(2k_q-k_B)
\nonumber \\
&&  ~~~~~~ \times [\bar{u}_{k_q,s} v_{-k_q,s}]^{{}^3P_0},
\label{overlap}
\end{eqnarray}

where, $k_q - k_B$ and $k_B -k_q$ represent the  momentum of heavy quark and anti-quark of $A$ meson
such that total momentum of meson $A$ is zero. The meson $C$ take a anti-quark of momentum $k_q - k_B$ of $A$ meson and one quark of momentum $k_q$ from $^3P_0$ pair such that its total momentum is $-k_B$. Similarly, the momentum of quark and anti-quark composing $B$ meson will be $k_q$ and $k_B - k_q$, respectively so that total momentum of $B$ meson is $k_B$. The term in the square bracket represents the wave function of $q\bar{q}$ pair produced in vacuum.
We take the harmonic oscillator potential for the bound state of wave function \cite{bonnaz,friman}. Considering the nodal structure of the wave function of the mesons $\phi$, we use the method of change of variable i.e., momentum in the following way, 
\begin{eqnarray}
k_q^\prime = k_q -{ 1+ r^2 \over 1 +2 r^2} k_B.
\label{change}
\end{eqnarray}

While doing so we allow the parent and daughter mesons to have a different sized wave functions by defining a variable $r$ = $\alpha/ \beta$, here $\alpha$ and $\beta$ represent the strengths of the initial and outgoing meson wave functions respectively \cite{ackleh, ei, vij}. The values of $r$ and $\beta$ are determined by fixing the experimental values of the partial decay width of $\psi(4040)$ to $DD$, $D D^*$ and $D^*D^*$ pairs and calculated values of $r$ and  $\beta$ are 1.04 and 0.3 GeV, respectively \cite{friman}.  

 
In the present problem for the different size of initial and final meson wave functions the resulting decay rate for the different charmonium decaying to $D \bar{D}$ rates of can be represented as \cite{friman}
\begin{align}
\Gamma(\psi(3686) \rightarrow D + \bar{D}) &=
\frac{\pi^{1/2} E_D ^2}{m_\psi(3686)} \gamma ^2 \frac{2^{7} (3+2r^2)^2 (1-3r^2)^2}{3^2(1+2r^2)^7}  \nonumber\\
 & \times x^3 ((1+ \frac{2r^2 (1+r^2)}{(1+2r^2)(3+2r^2)(1-3r^2)}x^2)^2  e^{-\frac{x^2}{2(1+2r^2)}},
\label{eq:p2}
\end{align}
\begin{align}
\Gamma(\psi(3770) \rightarrow D + \bar{D}) &=
 \frac{\pi^{1/2} E_D ^2}{m_\psi(3770)} \gamma ^2 \frac{2^{11} 5}{3^2} (\frac{r}{1+2r^2})^7 x^3 \nonumber\\
& \times (1- \frac{1+r^2}{5(1+2r^2)} x^2)^2 e^{-\frac{x^2}{2(1+2r^2)}}.
\label{eq:p3}
\end{align}
and
\begin{align}
\Gamma(\chi(3556) \rightarrow D + \bar{D}) &= \frac{\pi^{1/2} E_D ^2}{m_\psi(3556)} \gamma ^2 \frac{2^{10} r^5 (1+r^2)^2}{15(1+2r^2)^7} x^5 e^{-\frac{x^2}{2(1+2r^2)}}.
\label{eq:c1}
\end{align}
Here, $E_D^2$ = ${m^2_D} + {P^2_D}$,  momentum $P_D = \sqrt{\frac{(m^2_A - (m_B - m_C)^2)(m^2_A - (m_B + m_C)^2)}{2m_A}}$. The parameter $x$ is  expressed as,
 $x$ = $\frac{1}{\beta}$ $\sqrt{(m_A^2/4 - m_B^2)}$. In the above equations,  parameter $\gamma$ represents the strength of the $^3P_0$ vertex and calculated by fitting the experimental value of $\Gamma$($\psi(3770)$ $\rightarrow$ $D \bar{D}$)  and comes out to be 0.281 \cite{friman}. We will incorporate the medium modified mass of $D$ meson in the above equations and will calculate the decay width of charmonium states to $D \bar{D}$ pairs.

 \section{Results and Discussion}
 \label{sec_results_discussions}
In  \cref{sub_shift_mass_decay} we will discuss the result of our present work on the in-medium masses and decay constants of isospin averaged pseudoscalar $D(D^+,D^0)$ and $B(B^+, B^0)$ mesons in isospin asymmetric hot and dense strange hadronic matter. The effects of in-medium masses of $D(D^+, D^0)$ mesons on the  in-medium decay width of charmonium  $\psi(3686)$, $\psi(3770)$ and $\chi(3556)$ will be discussed in \cref{sub_shift_decaywidth}.
 In the present work we consider the nuclear matter saturation density as 0.15 fm$^{-3}$. Further, we take the average value of coupling constants $g_{{DN\Lambda_c}}$ $\approx$ $g_{{DN\Sigma_c}}$ $\approx$ $g_{{BN\Lambda_c}}$ $\approx$ $g_{{BN\Sigma_c}}$ $\approx$
 6.74 \cite{wang2}.
 On the hadronic side the values of masses of $M_{D^+}$,  $M_{D^0}$, $M_{B^+}$ and $M_{B^0}$ are taken as 1.869, 1.864, 5.279 and 5.280 GeV, respectively. The values of decay constants $f_{D}$ and $f_{B}$  for $D$ and $B$ mesons  are taken as 210 and 190 MeV, respectively. The continuum threshold parameters  $s_0$ for $D$ and $B$ mesons are taken as 6.2 and 33.5 GeV$^2$, respectively.

  We shall show the variation of mass shift and decay shift of $D$ and $B$ mesons as a function of squared Borel mass parameter, $M^2$.
   We chose the Borel window such that there is no variation in the mass or decay constant within that region of $M^2$. The Borel window chosen for the masses of  $D(D^+, D^0)$ and $B(B^+, B^0)$ mesons as (3.5-5.5) and (29-32) GeV$^2$, respectively, whereas for the decay constant of $D(D^+, D^0)$ and $B(B^+, B^0)$ mesons we chose the Borel window as  (2.5-4.5), and (25-29) GeV$^2$, respectively.

\begin{table}
\begin{tabular}{|l|l|l|l|l|l|l|}
\hline
   & & \multicolumn{4}{c|}{I=0}   \\
\cline{3-6}
&$f_s$ & \multicolumn{2}{c|}{T=0}& \multicolumn{2}{c|}{T=100 MeV} \\
\cline{3-6}
 & &  $\rho_0$ &$4\rho_0$&$\rho_0$&$4\rho_0$ \\ \hline
$\delta m_{D^{+}}$  &0&-64.3&-110&-55 &-103\\ \cline{2-6}
  & 0.5 &-77&-115 & -67 & -107 \\ \cline{1-6}
$\delta m_{D^{0}}$  &0 &-92 &-163 & -79 &-153 \\  \cline{2-6}
  &0.5 & -110 &-170 & -96&-158 \\  \hline
$\delta m_{B^{+}}$ &0&-672&-1270&-586 &-1203\\ \cline{2-6}
 & 0.5 &-795 &-1316 & -703 & -1240 \\ \cline{1-6}
$\delta m_{B^{0}}$  &0 &-480 &-904 & -418 &-857 \\  \cline{2-6}
 &0.5 & -568 &-938 & -502&-883 \\  \hline
\end{tabular}
\begin{tabular}{|l|l|l|l|l|l|l|}
\hline
   & & \multicolumn{4}{c|}{I=0.5}   \\
\cline{3-6}
&$f_s$ & \multicolumn{2}{c|}{T=0}& \multicolumn{2}{c|}{T=100 MeV} \\
\cline{3-6}
 & &  $\rho_0$ &$4\rho_0$&$\rho_0$&$4\rho_0$ \\ \hline
$\delta m_{D^{+}}$  &0&-68&-112&-60 &-108\\ \cline{2-6}
  & 0.5 &-84 &-126 & -73 & -132 \\ \cline{1-6}
$\delta m_{D^{0}}$  &0 &-82 &-142 & -72 &-137 \\  \cline{2-6}
  &0.5 & -89 &-141 & -79&-132 \\  \hline
$\delta m_{B^{+}}$ &0&-600&-1123&-535 &-1090\\ \cline{2-6}
 & 0.5 &-666 &-1133 & -595 & -1071 \\ \cline{1-6}
$\delta m_{B^{0}}$  &0 &-509 &-923 & -453 &-894 \\  \cline{2-6}
 &0.5 & -629 &-1024 & -557&-960 \\  \hline
\end{tabular}

\caption{In above we tabulate the shift in masses of $D^+$, $D^0$, $B^+$ and $B^0$ mesons (in MeV) 
for baryonic densities $\rho_0$ and $4\rho_0$. The in-medium mass shift are given for isospin asymmetric parameters $I$ = 0 and 0.5, strangeness fractions $f_s = 0$ and $0.5$ and temperatures $T = 0$ and $100$ MeV.}
\label{table_mass_db_pseudo}
\end{table}

\begin{table}
\begin{tabular}{|l|l|l|l|l|l|l|}
\hline
   & & \multicolumn{4}{c|}{I=0}   \\
\cline{3-6}
&$f_s$ & \multicolumn{2}{c|}{T=0}& \multicolumn{2}{c|}{T=100 MeV} \\
\cline{3-6}
 & &  $\rho_0$ &$4\rho_0$&$\rho_0$&$4\rho_0$ \\ \hline
$\delta f_{D^{+}}$  &0&-6&-8.8&-5 &-8\\ \cline{2-6}
  & 0.5 &-7 &-9 & -6& -8.4 \\ \cline{1-6}
$\delta f_{D^{0}}$  &0 &-9 &-14 & -7 &-13 \\  \cline{2-6}
  &0.5 & -11 &-15 & -9&-14 \\  \hline
$\delta f_{B^{+}}$ &0&-100&-189&-87 &-179\\ \cline{2-6}
 & 0.5 &-119 &-196 & -105 & -185 \\ \cline{1-6}
$\delta f_{B^{0}}$  &0 &-71 &-135 & -62 &-127 \\  \cline{2-6}
 &0.5 & -85 &-140 & -75&-131 \\  \hline
\end{tabular}
\begin{tabular}{|l|l|l|l|l|l|l|}
\hline
   & & \multicolumn{4}{c|}{I=0.5}   \\
\cline{3-6}
&$f_s$ & \multicolumn{2}{c|}{T=0}& \multicolumn{2}{c|}{T=100 MeV} \\
\cline{3-6}
 & &  $\rho_0$ &$4\rho_0$&$\rho_0$&$4\rho_0$ \\ \hline
$\delta f_{D^{+}}$  &0&-3.5&-6.6&-3 &-4.2\\ \cline{2-6}
  & 0.5 &-4.2 &-5 & -3.5& -5.4 \\ \cline{1-6}
$\delta f_{D^{0}}$  &0 &-8 &-12 & -7 &-11 \\  \cline{2-6}
  &0.5 & -9 &-12 & -8&-11 \\  \hline
$\delta f_{B^{+}}$ &0&-89&-167&-80 &-162\\ \cline{2-6}
 & 0.5 &-99 &-169 & -89 & -159 \\ \cline{1-6}
$\delta f_{B^{0}}$  &0 &-76 &-137 & -67 &-133 \\  \cline{2-6}
 &0.5 & -94 &-152 & -83&-143 \\  \hline
\end{tabular}

\caption{In above we tabulate the shift in decay constant of $D^+$, $D^0$, $B^+$ and $B^0$ mesons (in MeV)
for baryonic densities $\rho_0$ and $4\rho_0$. The in-medium  shift of decay constant are given for isospin asymmetric parameters $I$ = 0 and 0.5, strangeness fractions $f_s = 0$ and $0.5$ and temperatures $T = 0$ and $100$ MeV.}
\label{table_decay_db_pseudo}
\end{table}

\subsection{Shift in masses and decay constants}
\label{sub_shift_mass_decay}
\begin{figure}
\centering
\includegraphics[width=16cm,,height=14cm]{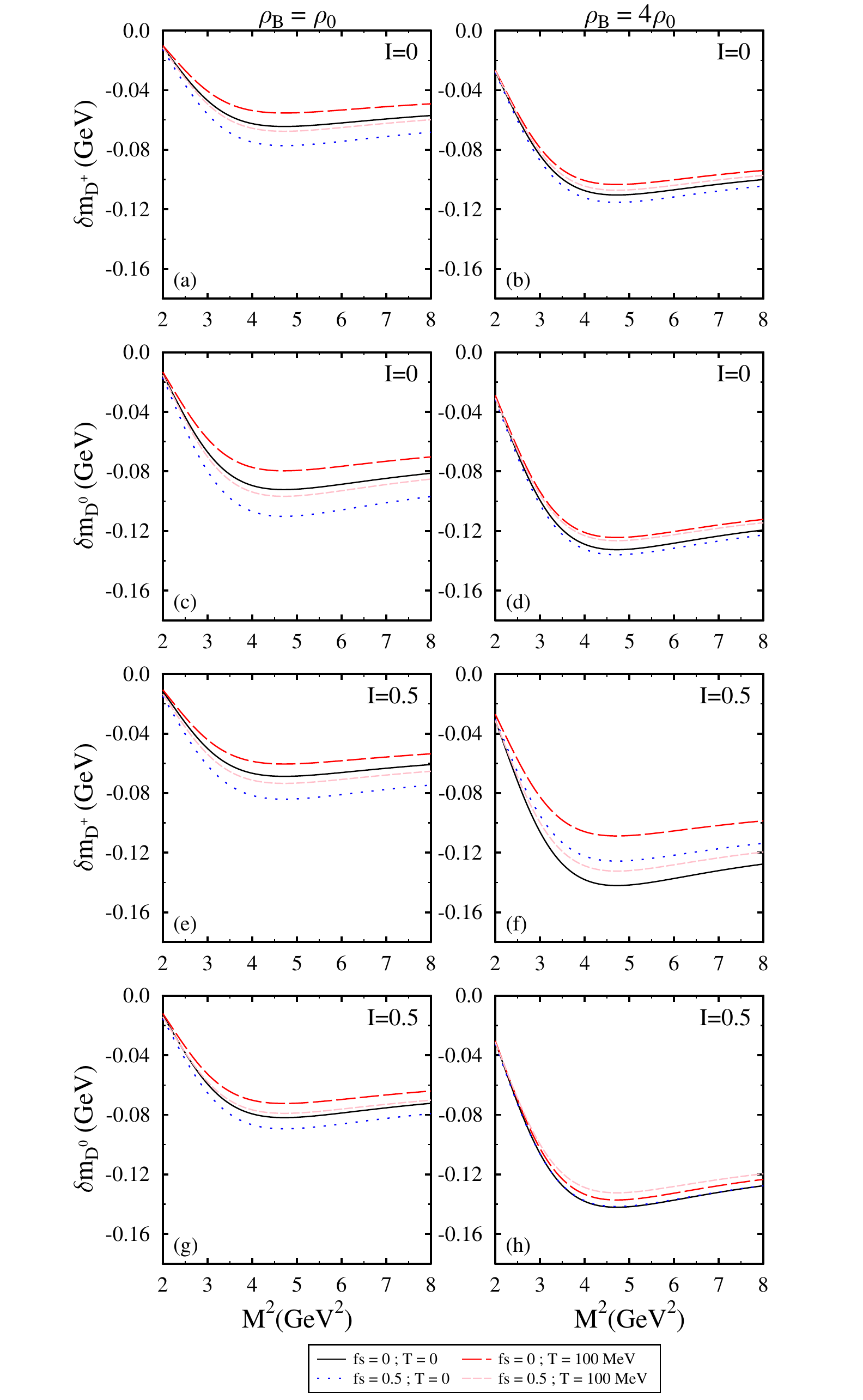}
\caption{Figure shows the variation of mass shift of pseudoscalar $D^+$ and $D^{0}$ mesons as a function of squared Borel mass parameter, $M^2$ for baryonic densities $\rho_0$ and $4\rho_0$. The results are given for isospin asymmetric parameter $I = 0$ and $0.5$, temperatures $T = 0$ and 100 MeV and strangeness fractions $f_s = 0$ and $0.5$.}
\label{dmass}
\end{figure}
\begin{figure}
\centering
\includegraphics[width=16cm,,height=14cm]{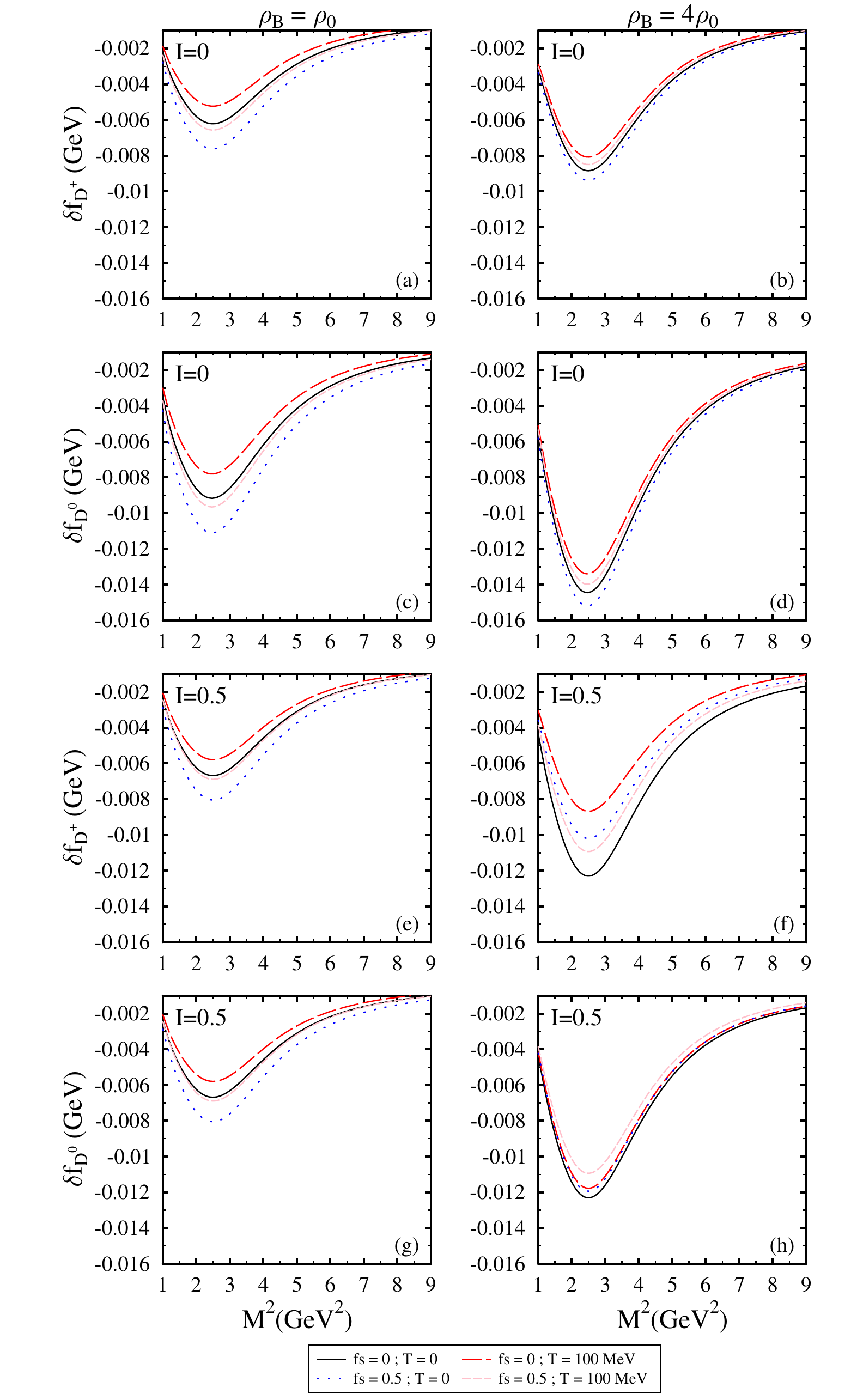}
\caption{Figure shows the variation of  shift in decay constant of pseudoscalar $D^+$ and $D^{0}$ mesons as a function of squared Borel mass parameter, $M^2$ for baryonic densities $\rho_0$ and $4\rho_0$. The results are given for isospin asymmetric parameter $I = 0$ and $0.5$, temperatures $T = 0$ and 100 MeV and strangeness fractions $f_s = 0$ and $0.5$.}\label{ddecay}
\end{figure}

\begin{figure}
\centering
\includegraphics[width=16cm,,height=14cm]{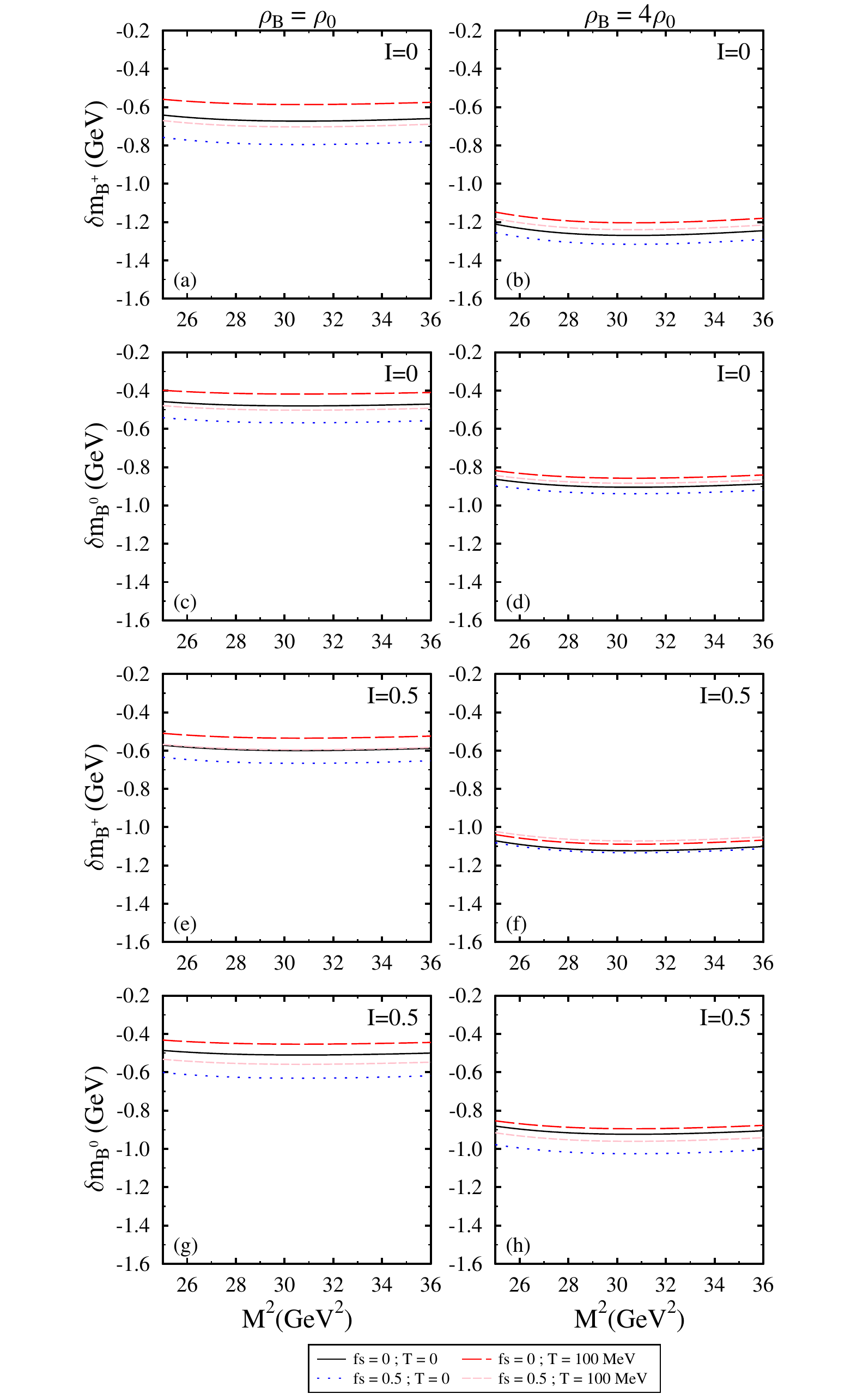}
\caption{Figure shows the variation of mass shift of pseudoscalar $B^+$ and $B^{0}$ mesons as a function of squared Borel mass parameter, $M^2$ for baryonic densities $\rho_0$ and $4\rho_0$. The results are given for isospin asymmetric parameter $I = 0$ and $0.5$, temperatures $T = 0$ and 100 MeV and strangeness fractions $f_s = 0$ and $0.5$.}\label{bmass}
\end{figure}

\begin{figure}
\centering
\includegraphics[width=16cm,,height=14cm]{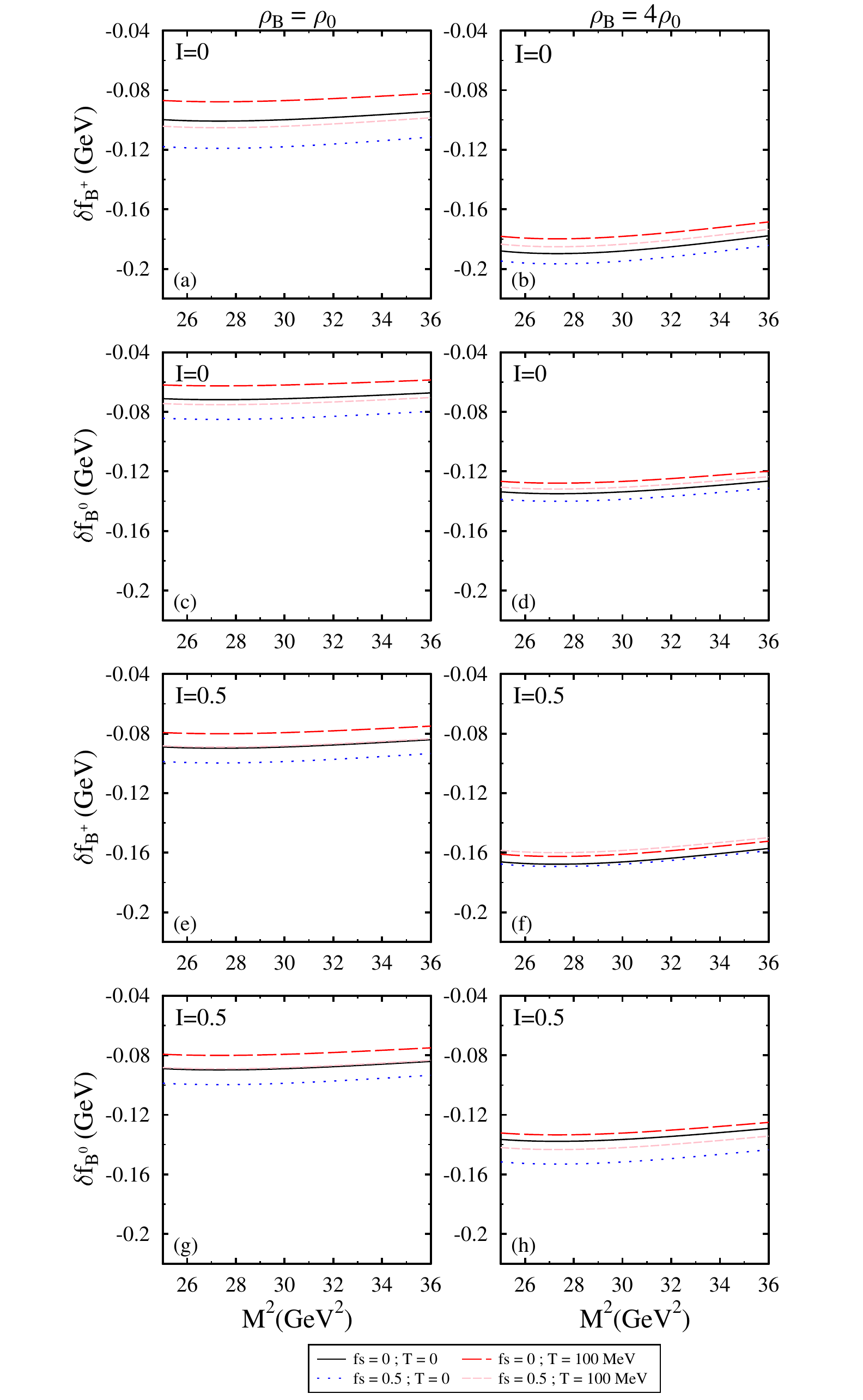}
\caption{Figure shows the variation of shift in decay constant of pseudoscalar $B^+$ and $B^{0}$ mesons as a function of squared Borel mass parameter, $M^2$ for baryonic densities $\rho_0$ and $4\rho_0$. The results are given for isospin asymmetric parameter $I = 0$ and $0.5$, temperatures $T = 0$ and 100 MeV and strangeness fractions $f_s = 0$ and $0.5$.}\label{bdecay}
\end{figure}

 
In \cref{dmass} (\cref{ddecay}) we plot the shift in mass (decay constant) of $D$ meson whereas in \cref{bmass} (\cref{bdecay}) we plot the shift in mass (decay constant) of $B$ mesons as a function of square of Borel mass parameter i.e., $M^2$. We represent the results of masses and decay shifts for the  baryonic densities $\rho_0$ and 4$\rho_0$. For each value of density, the results are shown for strangeness fractions $f_s$ = 0 and 0.5, and isospin asymmetric parameters, $I$ = 0 and  0.5. We compare the results at temperatures $T$ = 0 and  100 MeV. In \cref{table_mass_db_pseudo,table_decay_db_pseudo}, we tabulate the values of shift in masses and decay constants of $D$ and $B$ mesons for different conditions in the medium.

For a given value of temperature $T$, isospin asymmetric parameter $I$ and strangeness fractions $f_s$, the masses and decays constants of pseudoscalar $D$ and $B$ mesons are observed to decrease as a function of 
density of hadronic  medium. For example, as can be seen from \cref{table_mass_db_pseudo}, in symmetric nuclear matter, at $\rho_B = \rho_0$ and $T = 0$, the values of masses  of $D^{+}$,  $D^{0}$ $B^{+}$,  $B^{0}$ mesons decrease by $3.4\%$, $4.9\%$, $12.7\%$ and $9\%$, respectively from the vacuum value. From \cref{table_mass_db_pseudo} we note that the masses of $D^{+}$ and $D^{0}$  mesons are different in the isospin symmetric medium. This is because in the present work we consider the different masses of $u$ and $d$ quarks and this lead to different values of scalar quark condensates $\bar{u}u$ and $\bar{d}d$ and hence, different masses of $D^{+}$ and $D^{0}$  mesons in symmetric medium.

 The isospin asymmetry of the medium causes the mass splitting between $D^+$ and $D^0$, and  $B^+$ and $B^0$ mesons.
When we move from the isospin symmetric medium ($I = 0$) to isospin
asymmetric medium  ($I = 0.5$), the mass of $D^0$ and $B^+$
meson increases, whereas that of $D^+$ and  $B^0$ mesons decreases. For example, in nuclear medium at $\rho_B = \rho_0$ and $T = 0$, as we move from $I = 0$ to $0.5$, the in-medium mass of $D^0$ increases by 
$0.56 \%$ whereas the in- medium mass of $D^+$ decreases by $0.21 \%$.
The opposite behavior of  $D^0$ and $D^+$ mesons as a function of isospin asymmetry of the medium can be understood
in terms of the behavior of condensates $\left\langle u \bar{u} \right\rangle$ and $\left\langle d \bar{d} \right\rangle$.
Recall that the $D^0$ and $D^+$ mesons contain $u$ and $d$ quark, respectively and thus, there medium modification is because of the condensates corresponding to these quarks.
As can be seen from \cref{qu,qd}, due to non-zero value of scalar-isovector field $\delta$ in isospin asymmetric matter,  condensates $\left\langle u \bar{u} \right\rangle$ and $\left\langle d \bar{d} \right\rangle$ behave oppositely in an isospin asymmetric matter.  For example, in nuclear medium, at $\rho_B = \rho_0$ and temperature, $T = 0$, as we change $I$ from $0$ to $0.5$, the value of $\frac{\left\langle u \bar{u} \right\rangle}{\left\langle u \bar{u} \right\rangle_0} $ increases by 6$\%$  of its vacuum value , whereas the value of $\frac{\left\langle d \bar{d} \right\rangle}{\left\langle d \bar{d} \right\rangle_0}$ decreases by 4$\%$. Here,  $\left\langle u \bar{u} \right\rangle_0$ and $\left\langle d \bar{d} \right\rangle_0$ denote their vacuum values and are calculated through chiral SU(3) model as -1.401 $\times$ 10$^{-2}$ GeV$^3$ and  -1.401 $\times$ 10$^{-2}$ GeV$^3$, respectively.
At finite value of strangeness fraction e.g., at $f_s$ = $0.5$, 
the magnitude of shift in mass of  $D^0$ and $B^+$
mesons decreases by $19\%$ and $16\%$, whereas magnitude of shift in mass of  $D^+$ and $B^0$ mesons increases  by $9\%$ and $11\%$, as one move from $I = 0$ to $0.5$.

 For a given value of isospin asymmetry of the medium, the increase in the value of strangeness fractions $f_s$ i.e, inclusion of more number hyperons in the medium, cause  more decrease in the values of the masses and decay constants of $D$ and $B$  mesons. For example, at baryonic density $\rho_B = \rho_0$, isospin asymmetric parameter $ I= 0$ and temperature $T = 0$, if we move from nuclear medium ($f_s$ = 0) to strange hadronic medium ($f_s$ = 0.5), the percentage decrease in the mass (decay constant) of $D$ meson is observed to be 0.6$\%$ (0.2$\%$), whereas for $B$ meson these values shift to 2$\%$ (14$\%$), respectively. 
Note that the change in mass and decay constant of $D$ and $B$ mesons is more sensitive to isospin asymmetry of the medium as compared to the strangeness fraction. This is because the $D$ and $B$ meson we are investigating in the present work contain light $u$ or $d$ quark. The condensates corresponding to these quarks depend upon the scalar field $\sigma$ which in turn is more sensitive to isospin asymmetry of the medium as compared to strangeness fraction.  On the other side, the strange isoscalar field $\zeta$, which have strange quark content, changes appreciably as a functions of strangeness fraction of the medium as compared to isospin asymmetry. 
For example, at temperature $T = 0$ and isospin symmetric parameter $I = 0$, as we move from $f_s$ = 0 to  0.5, the value of $\zeta$ field changes by 2.7$\%$ and 14$\%$  at baryonic densities $\rho_ B = \rho_0$ and $\rho_B = 4\rho_0$, respectively, whereas for $\sigma$ field observed changes are nearly 3.5$\%$ and $1.6\%$, respectively. However, for fix value of strangeness fraction say $f_s$ = $0$, at $\rho_B = \rho_0$ and $T = 0$, as $I$ is changed from $0$ to $0.5$, the values of $\sigma$ and $\zeta$ changes by $1.3 \%$ and $0.17\%$, respectively at $\rho_0$, whereas at $4\rho_0$ these values shift to $10\%$ and $0.36\%$, respectively.

At finite baryonic density, the finite temperature of the medium is observed to cause an increase in the values of masses and decay constants of pseudoscalar $D$ and $B$ mesons. For example, in symmetric nuclear matter, at density $\rho_0$, the percentage increase in the values of the masses of $D^+$ and $D^0$ ($B^+$ and $B^0$) mesons are observed to be 0.5$\%$ and 0.7$\%$ (1.6$\%$ and 1.2$\%$), whereas, percentage increase in decay constants are observed to be 0.4$\%$ and 0.9$\%$ (6.8$\%$ and $4.7\%$), respectively,  as we increase the temperature of the medium from 0 to 100 MeV. This happens  because, at finite density, the drop in the $\sigma$ and $\zeta$ fields from vacuum value is less at finite temperature of the medium as compared to zero temperature and therefore, the magnitude of $\sigma$ and $\zeta$ fields will increase as one move from zero to finite temperature of the medium. For example, in
symmetric nuclear matter, at baryonic density $\rho_B$ = $\rho_0$, with
 change of temperature from $T$ = 0 to  100 MeV, the
magnitude of scalar field $\sigma$ increase by
7.1$\%$, whereas for $\zeta$ field, the increase in magnitude is about 1$\%$ which is consistent with the results of Ref. 
 \cite{pwang} in which the chiral quark mean field model was used to compute the in-medium properties of baryons through the scalar fields $\sigma$ and $\zeta$. The above changes in the values of scalar fields are further reflected in the quark and gluon condensates and which in turn modify the properties of $D$ and $B$ mesons as a functions of temperature.
 
 
 The results on the in-medium masses and decay constants of $D$ and $B$ mesons discussed above consider the contribution of next to leading order (NLO) term to the scalar quark condensate $\bar{q}q$ \cite{wang2}. However, if we consider only the leading order (LO) term, then there will be less drop in the masses and decay constants of $D$ and $B$ mesons as can be see from \cref{table_mass_indiv_db_pseudo_1,table_mass_indiv_db_pseudo_2}. In these tables we also tabulate the values of mass shift and decay shift considering the contribution of  scalar light quark condensates $\left\langle  \bar{q}q \right\rangle$ only. We observe that the contribution of all other condensates to in-medium properties is very small as compared to $\left\langle  \bar{q}q \right\rangle $. 
  For example, in symmetric nuclear medium, at $\rho_B=\rho_0$ and  $T = 0$, if we keep $\left\langle q \bar{q} \right\rangle $ only in the operator product expansion side of QCD sum rules, and put all the other condensates equal to zero then the shift in mass of $D^{+}$ meson changes by 11$\%$ to its original value. Also, in the present work, except the condensate $\langle q^\dag i D_0q\rangle_N$ , the  all other condensates are evaluated within the chiral SU(3) model. However, this condensate has little effect on medium modification of $D$ and $B$ mesons.  For example, in symmetric nuclear matter, at $\rho_B = \rho_0$ and $T = 0$, if we consider $\langle q^\dag i D_0q\rangle_N$ = 0, then value of mass shift of $D^+$ mesons changes by $3.7\%$ as compared to the situation when we consider all condensates.

 \begin{table}
\begin{tabular}{|l|l|l|l|l|l|l|l|l|l|l|}
\hline
& & \multicolumn{4}{c|}{I=0} & \multicolumn{4} {c|}{I=0.5}\\  \cline{3-10}

$D^+$& & \multicolumn{2}{c|}{T=0} &\multicolumn{2}{c|}{T=100 MeV}& \multicolumn{2}{c|}{T=0} & \multicolumn{2}{c|}{T=100 MeV}    \\\cline{3-10}

& & $\rho_0$ & $4\rho_0$  &$\rho_0$&4$\rho_0$&$\rho_0$&4$\rho_0$ & $\rho_0$ & 4$\rho_0$ \\ \hline
All Condensates  &NLO&-64.4&-110&-55 &-103 &-68 &-112 & -60 & -108\\ \cline{2-10}
  & LO &-45 &-74 & -38& -69 & -48& -76 & -42 & -73 \\ \cline{1-10}
$\left\langle d\bar{d} \right\rangle _N$ $\neq$ 0  &NLO &-57 &-91 & -49 &-84 &-62 &-93 &-54 &-90\\  \cline{2-10}
  &LO & -39 &-55 & -32&-52 &-42 &-57 &-36 &-54\\  \hline
$\langle \bar{q} i D_0 q\rangle _N$ =0 &NLO&-62&-101&-53 &-94 &-66 & -104 & -58 &-99\\ \cline{2-10}
 & LO &-43 &-65 & -36 & -60 &-46  &-67  &-40 &-64\\ \hline

\end{tabular}
\caption{In the above table mass shift of $D^{+}$ mesons (in MeV) are compared by considering the contribution of individual condensates.}
\label{table_mass_indiv_db_pseudo_1}
\end{table}

\begin{table}
\begin{tabular}{|l|l|l|l|l|l|l|l|l|l|l|}
\hline
& & \multicolumn{4}{c|}{I=0} & \multicolumn{4} {c|}{I=0.5}\\  \cline{3-10}

$D^0$& & \multicolumn{2}{c|}{T=0} &\multicolumn{2}{c|}{T=100 MeV}& \multicolumn{2}{c|}{T=0} & \multicolumn{2}{c|}{T=100 MeV}    \\\cline{3-10}

& & $\rho_0$ & $4\rho_0$  &$\rho_0$&4$\rho_0$&$\rho_0$&4$\rho_0$ & $\rho_0$ & 4$\rho_0$ \\ \hline
All Condensates  &NLO&-92&-163&-79 &-153 &-81 &-141 & -72 & -137\\ \cline{2-10}
  & LO &-65 &-113 & -56& -106 & -58 & -97 & -51& -93 \\ \cline{1-10}
$\left\langle d\bar{d} \right\rangle _N$ $\neq$ 0  &NLO &-84 &-141 & -72 &-132 &-74 &-121 &-65 &-117\\  \cline{2-10}
  &LO & -58 &-91 & -49&-85 &-51 &-77 &-44 &-73\\  \hline
$\langle \bar{q} i D_0 q\rangle _N$ =0 &NLO&-90&-154&-77 &-144 &-79 & -133 & -69 &-128\\ \cline{2-10}
 & LO &-63 &-104 & -54 & -97 &-56  &-88  &-48 &-85\\ \hline

\end{tabular}
\caption{In the above table mass shift of $D^{0}$ mesons (in MeV) are compared by considering the contribution of individual condensates.}
\label{table_mass_indiv_db_pseudo_2}
\end{table}
 
 In literature, using quark meson coupling model, the shift in masses of $D^+$ and $D^-$ mesons were observed to be nearly -60 MeV \cite{tsushima16}.
 By considering the vector mean field contribution, the shift in mass of $D^+$ meson of about -140 MeV  and repulsive potential of about +20 MeV for $D^-$ meson was observed at nuclear matter density in Ref. \cite{sibi17}.  
 In Ref. \cite{jim} using coupled channel approach with $t-$channel vector-exchange model, authors predicted positive (negative) mass shift of $D^-$($D^+$) meson of around +32(-27) MeV, respectively at density $\rho_0$.  On the other hand, using QCD sum rule and maximum entropy method authors observed the shift in mass of $D^+$($D^-$) mesons of about 23 (38) MeV \cite{suzu}, at nuclear saturation density.  
  In Ref. \cite{haya}, shift in mass of $D$ meson had been studied using QCD sum rules in symmetric nuclear matter at zero temperature  and the observed mass shift was -65 MeV.
In Ref. \cite{hilger} authors found strong splitting between of about 60 MeV between $D^{+}$ and $D^{-}$ mesons  by expressing the  correlation function using the Lehmann representation and separating the odd and even part of QCD sum rule . However, in the present work we calculate the average shift in mass by dividing the correlation function into vacuum and medium part \cite{wang2,haya}. 
  In \cite{wang2}, using the QCD sum rules the shift in masses and decay constants of $D$ and $B$ mesons have been obtained upto leading order term and next to leading order term in symmetric nuclear matter at zero temperature. At $\rho_B = \rho_0$, the values of shift in masses and decay constants of $D$($B$) mesons upto next to leading order term were observed to be -72(-473) and -6(-71) MeV,  whereas upto leading order terms these values changes to -47(-329) and -4(-48) MeV, respectively.  In all of above references the properties of $D$ and $B$ mesons had been studied  only in nuclear medium. 
 
\begin{figure}
\centering
\includegraphics[width=12cm,,height=10cm]{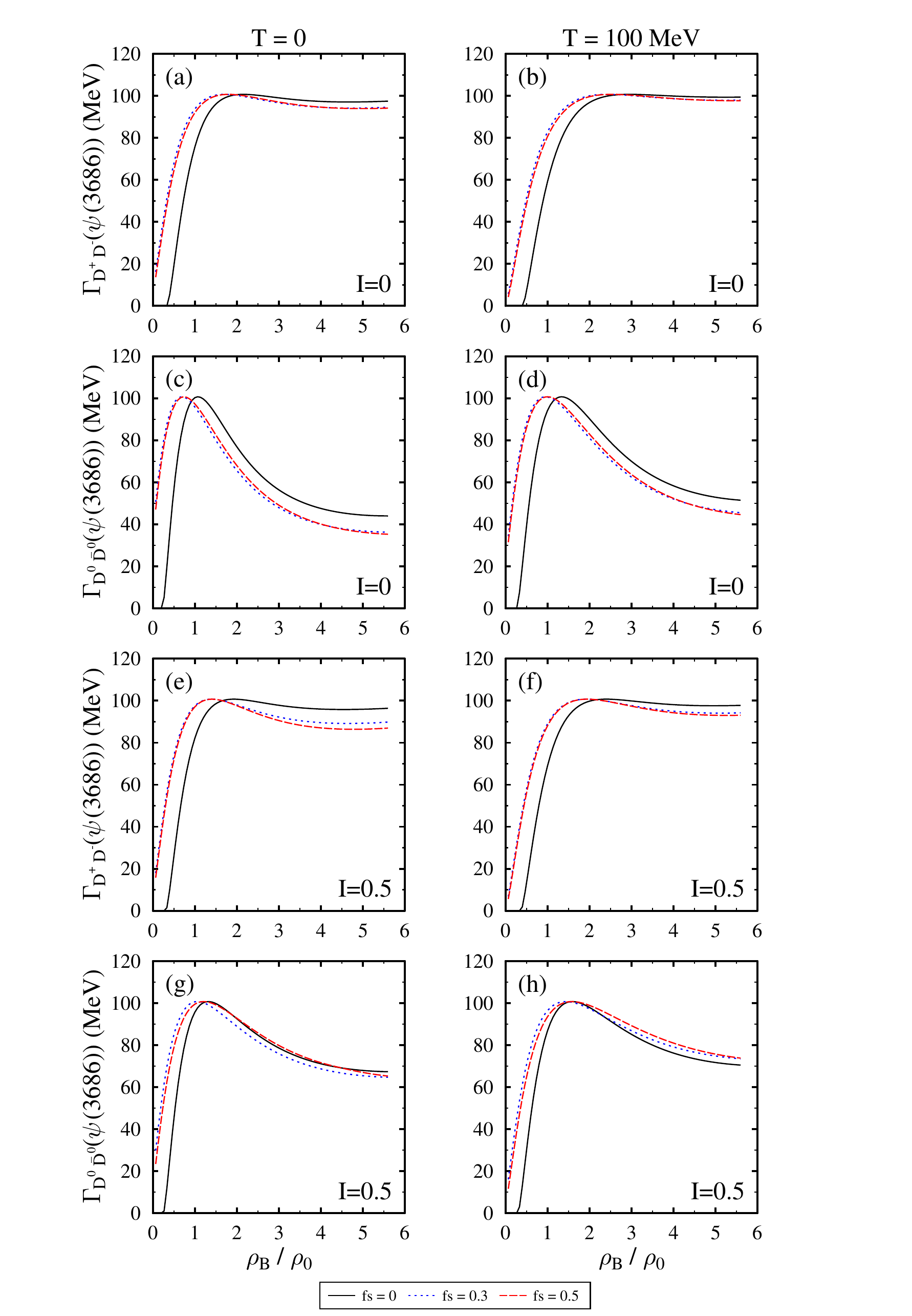}
\caption{Figure shows the variation of in-medium decay width of charmonium $\psi(3686)$ to $D\bar{D}$ pairs as a function of baryonic density $\rho_B$ (in units of nuclear saturation density $\rho_0$). The results are given for isospin asymmetric parameter $I = 0$ and $0.5$, temperatures $T = 0$ and 100 MeV and strangeness fractions $f_s = 0$ and $0.5$.}
\label{fig_psi3686}
\end{figure}

\begin{figure}
\centering
\includegraphics[width=12cm,,height=10cm]{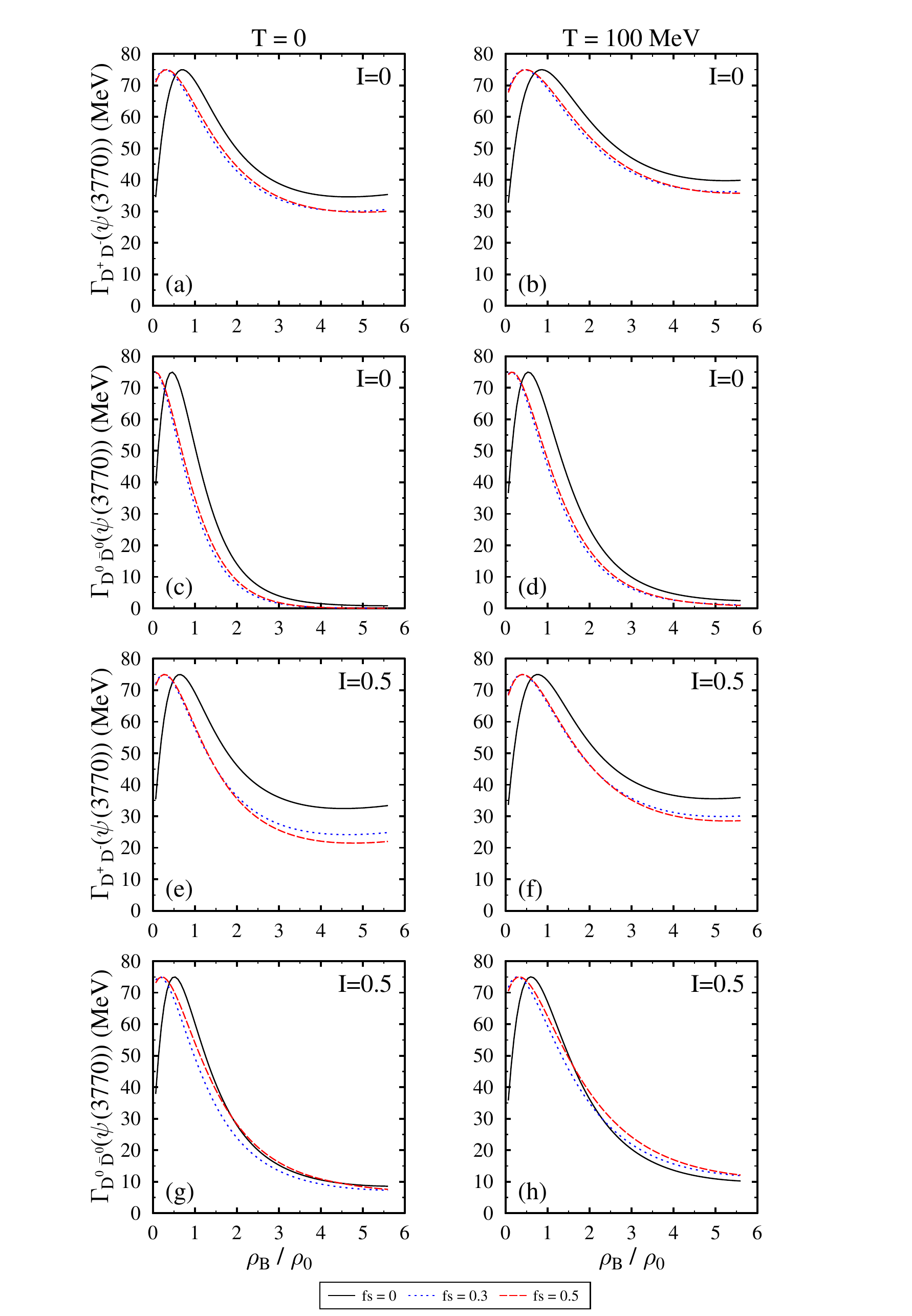}
\caption{Figure shows the variation of in-medium decay width of charmonium $\psi(3770)$ to $D\bar{D}$ pairs as a function of baryonic density $\rho_B$ (in units of nuclear saturation density $\rho_0$). The results are given for isospin asymmetric parameter $I = 0$ and $0.5$, temperatures $T = 0$ and 100 MeV and strangeness fractions $f_s = 0$ and $0.5$.}
\label{fig_psi3770}
\end{figure}

\begin{figure}
\centering
\includegraphics[width=12cm,,height=10cm]{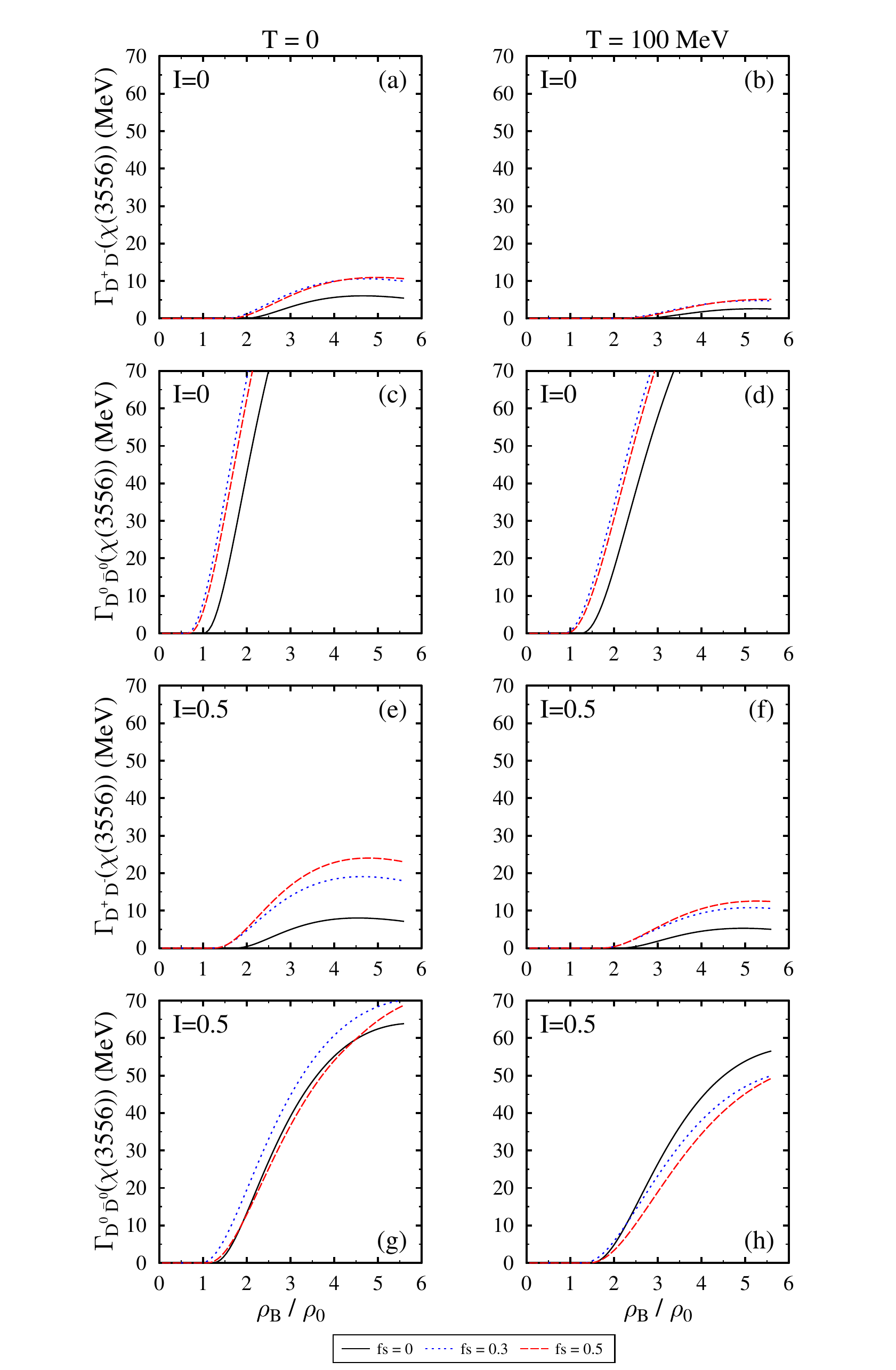}
\caption{Figure shows the variation of in-medium decay width of charmonium $\chi(3556)$ to $D\bar{D}$ pairs as a function of baryonic density $\rho_B$ (in units of nuclear saturation density $\rho_0$). The results are given for isospin asymmetric parameter $I = 0$ and $0.5$, temperatures $T = 0$ and 100 MeV and strangeness fractions $f_s = 0$ and $0.5$.}
\label{fig_chi3556}
\end{figure}

\subsection{Shift in decay width}
\label{sub_shift_decaywidth}
Now we investigate the effects of above discussed medium modified masses of $D$ meson on the in-medium decay width of the charmonium  $\psi(3686)$, $\psi(3770)$ and $\chi(3556)$  to $D\bar{D}$ pairs.
In the present investigation, we  neglect the medium modification of higher charmonium states, however, at last we shall also discuss the implications of medium modification of these states.  

\begin{table}
\begin{tabular}{|l|l|l|l|l|l|l|l|l|l|l|}
\hline
   & & \multicolumn{4}{c|}{I=0} & \multicolumn{4}{c|}{I=0.5}  \\
\cline{3-10}
& & \multicolumn{2}{c|}{T=0} & \multicolumn{2}{c|}{T=100 MeV} &\multicolumn{2}{c|}{T=0} & \multicolumn{2}{c|}{T=100 MeV}\\
\cline{3-10}
 & $f_s$&  $\rho_0$ &$4\rho_0$&$\rho_0$&$4\rho_0$ &   $\rho_0$ & $4\rho_0 $ & $\rho_0$ & $4\rho_0$\\ \hline
$\Gamma_{D^+{D^-}}(\psi(3686))$ &0&75&97&59 &99.9 &82 &96 & 69  & 98\\ 
 & 0.5 &92 &94 & 80 & 98.7 &97 & 87 & 88 & 94 \\ \cline{2-10} 
$\Gamma_{D^0 \bar{D^0}}(\psi(3686))$  &0 &100 &47 & 94 &58&95&71 &86 &76 \\  
 &0.5 & 97 & 40&100  & 52&99 &71 &93 &80 \\ \cline{2-10} 
 $\Gamma_{D^+ {D^-}}(\psi(3770))$ &0 &71 &35 & 74 &41 &69 &32 &73 &36 \\   
 &0.5 &  63 &31 & 69&38 &58 &22 &66 &30 \\ \cline{2-10} 
 $\Gamma_{D^0 \bar{D^0}}(\psi(3770))$ &0 &51 &1.3 & 62 &4.6 &60 &10 &67 &13.6 \\   
 &0.5 & 35  & 0.3 & 47& 2.7 &54 &10 &62 &17 \\  \cline{2-10}
 $\Gamma_{D^+{D^-}}(\chi(3556))$  &0&0&5.6&0 &1.7 &0&7&0&4\\ 
  & 0.5 &0&9.8 &0 & 3.5 &0 & 23 & 0 & 11 \\ \cline{2-10}
$\Gamma_{D^0{\bar{D^0}}}(\chi(3556))$  &0 &0 &116 & 0 &84 &0 &55 &0 &44 \\  
 &0.5&5.6 &140 &0.2 & 102&0 &54 &0 &34\\  \hline
\end{tabular}

\caption{In the above we tabulate the values of in-medium decay width of charmonium $\psi(3686)$, $\psi(3770)$ and $\chi(3556)$ to $D\bar{D}$ pairs (in MeV) for different conditions of the medium.}
\end{table}
\Cref{fig_psi3686} shows the variation of partial decay width of  $\psi(3686)$ state decaying into $D \bar{D}$ mesons as a function of density of the hadronic medium. 
 We represent the decay width  for the strangeness fractions $f_s$ = 0 and 0.5, isospin asymmetric parameter $I$ = 0 and 0.5, and temperatures $T$ = 0 and 100 MeV.
 For a given value of temperature and isospin asymmetry of the medium, initially for very low baryonic density  the decay width of charmonium $\psi(3686)$ to $D\bar{D}$ pairs remain zero and then start increasing with increase in the density of the medium, upto certain value of density beyond which it start decreasing with further increase in the density. The zero value of decay width at low baryonic density is because the threshold value of $D\bar{D}$ pair is more than the mass of $\psi(3686)$. However, with increase in density, the mass of $D$ mesons start decreasing and the  in-medium threshold value of $D\bar{D}$ pair falls below the mass of $\psi(3686)$. This result in the finite decay width of  $\psi(3686)$ to $D\bar{D}$ pair. 
 The initial decrease of $D$ meson mass with density causes an increase in decay width  of $\psi(3686)$ to $D\bar{D}$ pairs upto certain density. 
 However, at higher values of density, the much decrease of the mass of $D$ meson start causing decrease in the decay width.
  For example, at $T = 0$ MeV, in the symmetric nuclear matter, the decay width of $\psi(3686)$ to $D^+ D^-$ remain zero upto density $0.33 \rho_0$, then increases with density, attaining maximum value $100.6$ MeV  at $2.13 \rho_0$ and beyond this it start decreasing with further increase in density. Note that at density $2.13 \rho_0$, the mass of $D^+$ meson undergoes a shift of  $-96.62$ MeV. The above discussed behaviour of decay width of $\psi(3686)$ as function of density can be understood from the expression of its decay width given in \cref{eq:p2}. Note that the expression of decay width is the product of  polynomial and Gaussian parts.
  As a function of density of the medium (considering medium modified $D$ meson masses), the polynomial part of the expression, which first increase and then decrease as a function of density of the medium,  dominate over the Gaussian part. This result in the above discussed trend of in-medium  decay width of $\psi(3686)$ as a function of baryonic density. 
 
  When we move from nuclear  to strange hadronic medium, the decay width of $\psi(3686)$ to $D^+ D^-$  increase with increase in $f_s$ below $2\rho_0$, whereas above $2\rho_0$, increase of $f_s$ causes decrease in the decay width.
This is because as a function of $f_s$ the mass of  $D^+$ meson decreases and above $2\rho_0$ the decrease is so much that nodal structure of wave-function comes into play and this causes more decrease in the decay width at finite $f_s$ as compared to $f_s = 0$.  For example, at $4\rho_0$, for $f_s = 0.5$ and $T = 0$ MeV,  the shift in the mass of $D^{+}$ is $-115.38$ MeV which is more in magnitude than at $f_s = 0$, having value $-110.41$ MeV and therefore, the decay width of
$\psi(3686)$ to $D^+ D^-$ will be less at $f_s = 0.5$ (having value $94.66$ MeV) as compared to $f_s = 0$ (having value 97.39 MeV).
The in-medium mass of $D$  mesons increases with increase in the temperature of the medium and this  causes the opposite
 behavior of decay width of  $\psi(3686)$ as a function of temperature as compared  to the behaviour as a function strangeness fraction of the medium. 
This means below certain density the decay width of $\psi(3686)$
will decrease with increase in the temperature of the medium,  whereas above that it will increase with further increase of temperature.
For example, in symmetric nuclear matter,  below density $2.5 \rho_0$, the decay width decrease as we move from $T = 0$ to $T = 100$ MeV, whereas above this density the decay width increase as a function of $T$.
At temperature $T = 100$ MeV and baryonic density $\rho_B = \rho_0$ ($4\rho_0$), the decay width of $\psi(3686)$ to $D^+ D^-$ is observed to be $59.11$ (99.95)  MeV which is less (more) than value $75.69$ (97.39) MeV,  at $T = 0$ MeV.  
 Since the $D^0$ mesons undergo large mass drop
as compared to the $D^{+}$ mesons and therefore, the decay  of $\psi(3686)$ to $D^0 \bar{D^0}$  starts at less value of density as compared to what it was for $D^+ D^-$ pair. Large decrease of $D^{0}$ mass as compared to $D^{+}$ meson also causes the turn over of the behavior of decay width from increasing to decreasing trend at lower density. For example, in asymmetric nuclear matter with $I = 0.5$,
and at $T = 0$ MeV, beyond density $1.33 \rho_0$ the decay width  of $\psi(3686)$ to $D^0 \bar{D^0}$ start decreasing with increase in density whereas for the decay of $\psi(3686)$ to $D^+ D^-$ this value of density changes to $2\rho_0$.

In \cref{fig_psi3770} we show the in-medium value of $\Gamma_{D\bar{D}}(\psi(3770))$ as a function of baryon density of the medium. 
Since the vacuum threshold value of $D\bar{D}$ is larger than the  
mass of $\psi(3770)$ and therefore, $\Gamma_{D\bar{D}}(\psi(3770))$
have finite value even at zero baryon density. Similar to $\psi(3686)$, the in-medium decay width of $\psi(3770)$ first increase with increase in the density of the medium, reaching at some peak value and then start decreasing with the further increase
in the density. At temperature $T = 0(100)$ MeV, in the symmetric hadronic matter (I = 0), for strangeness $f_s = 0$ and $0.5$, the decay width of $\psi(3770)$ to $D^{+}D^{-}$ increase upto density $0.73 \rho_0$ ($0.866 \rho_0$) and $0.33\rho_0 (0.47\rho_0)$, respectively  
and then start decreasing.
 In the asymmetric matter with $I = 0.5$, the above values of densities at which turn over take place shift to $0.67\rho_0$ ($0.8 \rho_0$) and $0.27\rho_0$ ($0.4\rho_0$) at $f_s$ = 0 and 0.5, respectively. The maximum value of decay width $\Gamma_{D\bar{D}}(\psi(3770))$ is $75$ MeV. The decay width of $\psi(3770)$ decreases more rapidly as a function of density as compared to $\psi(3686)$. For example, in symmetric nuclear matter at $T = 0$ MeV and $\rho_B$ = $4\rho_0$, the decay width of $\psi(3686)$  and $\psi(3770)$ to $D^{+}D^{-}$ pairs decrease by $3.2\%$ and 40 $\%$ from their peak value. For the decay of $\psi(3686)$  and $\psi(3770)$ to $D^{0}\bar{D^{0}}$ pairs the above values of percentage change shift to $52.75\%$ and $98.01\%$, respectively.
 
 From \cref{fig_psi3770}, we observe that in isospin symmetric strange hadronic matter at zero temperature, the decay width of $\psi(3770)$ to $D^{0}\bar{D^{0}}$ pairs become almost zero above baryonic density $5\rho_0$. However, the zero decay width at higer densities is not observed for other conditions of the  medium.
 Note that in the present calculations using chiral SU(3) model and QCD sum rule, at density $5\rho_0$, in symmetric matter with $f_s = 0.5$ and $T = 0$ MeV,  the in-medium mass of $D^{0}$ mesons decreases by 174 MeV. This much of drop in the masses of $D$ mesons is not observed for any other values of different parameters and hence, zero value of decay width is not observed for other conditions in the present work.

From the above discussed results on charmonium decay width in nuclear medium we observe that the  internal structure of $D$ mesons play an important role \cite{friman} because of which the decay width first increase with density and then start decreasing. 
Mere the level crossing of $D\bar{D}$ threshold i.e., the decrease of in-medium mass of  $D\bar{D}$ pair below the mass of $\psi(3686)$ or $\psi(3770)$ does not guarantee for the decay of charmonium to $D\bar{D}$ pairs and hence, the decrease in the yield of $J/\psi$ mesons due to decay of higher charmonium states.
However, if one does not take into account the internal 
structure of $D$ mesons and consider them  as a point particle then the expression of partial decay width for $\psi(3770)$ can be expressed as
 \begin{eqnarray}
\Gamma_{D\bar{D}}(\psi(3770)) = \frac{g_{\psi DD}^2}{24\pi} \frac{(m_\psi^2 - 4{m}^2_D)^{3/2}}{{m}^2_\psi},
\label{slength}
\end{eqnarray}
where the coupling constant $g_{\psi DD}^2$ has the value 15.4 \cite{friman}.
The values of $\Gamma_{D \bar{D}}(\psi(3770))$ calculated using above expression will be very large as compared to the calculations of $^3P_0$ model due to available phase space. For example, in symmetric nuclear medium, at $T = 0$ MeV and density 2$\rho_0$,  we observe the decay width as 540 MeV, whereas at  finite temperature ($T=100$ MeV), this value reduces to 489 MeV.

In \cref{fig_chi3556}, we plot the  in-medium decay  width of $\chi (3556)$ state into $D \bar{D}$ pairs as a function of density of the hadronic medium.
  We observe that above certain value of baryonic density,  the partial decay width of $\chi(3556)$ state increases as a function of density of medium. 
 The zero value of $\Gamma_{D \bar{D}}(\chi(3556))$, at low baryonic density  is because the mass of  $\chi(3556)$ state is less than the total in-medium mass of $D\bar{D}$ pairs
 and therefore, the decay of $\chi(3556)$ to $D\bar{D}$ pairs is not possible. 
 As discussed earlier, the increase in the density of the medium cause a decrease in the mass of $D$ meson and therefore it results in the significant increase in the partial decay width of $\chi(3556)$ into $D \bar{D}$ pairs as a function of baryon density beyond a certain value of baryon density. Unlike, $\psi(3686)$ and $\psi(3770)$, we do not observe decrease in the decay width of $\chi(3556)$ at higher densities. 
 This is because the polynomial part of the decay width of $\chi(3556)$ always increases with the increase in the density of the medium unlike $\psi(3686)$ and $\psi(3770)$ for which polynomial part of the decay width first increase with density and then start decreasing.
  For a constant value of temperature, density and isospin asymmetric parameter of the medium, the observed  partial decay width has slightly more value in the strange medium than the non-strange medium except for the decay of $\chi(3556)$ to $D^0 \bar{D^0}$ in asymmetric matter. This is because of decrease in the mass of $D$ mesons as a function of strangeness fraction. Hence, we can conclude that presence of hyperons  in the medium may facilitate the $J/\psi$ suppression.
Since the finite temperature of the medium 
 increase the mass of $D$ meson and therefore, this will result in the decrease of 
  partial decay width of $\chi(3556)$ state to $D \bar{D}$ pairs. 
As we discussed earlier, the isospin asymmetry of the medium causes the  mass splitting between the $D$ and $D ^0$ mesons and this effect is further reflected in the decays width of charmonium states. The decrease in the mass of $D^{+}$ mesons as a function of isospin asymmetry of the medium causes an increase in the decay width of $\chi(3556)$ to $D^+ D^-$, whereas for the decay  of $\chi(3556)$ to $D^0 \bar{D^0}$ the situation is opposite.

 In the above calculations, while calculating the decay width of charmonium states we have not considered the medium modification of charmonium masses. In \cite{arvind3}, the in-medium decay widths of charmonium states were calculated using medium modified masses of $D$ mesons and charmonium states and results were compared with the situation when charmonium modification was neglected.
The in-medium masses of $D$ mesons were calculated in \cite{arvind3} using the generalized chiral SU(4)
 model, whereas in our present work we used QCD sum rules for calculating the medium modified masses of $D$ mesons. At higher baryonic densities larger drop in the masses of $D$ mesons was observed  within SU(4) model as compared to the calculations using QCD sum rules of present work and this results in zero decay width of $\psi(3686)$ and $\psi(3770)$ at $4.5\rho_0$ and $2.8\rho_0$, without considering medium modifications of charmonium masses in \cite{arvind3}. However, no node was observed when the medium modified charmonium masses were considered along with modification of $D$ mesons \cite{arvind3}. In  \cite{friman} the charmonium decay width were calculated considering the medium modification of $D$ mesons and it was observed that the nodes in the decay of  $\psi(3686)$ and $\psi(3770)$ will be observed when the $D$ mesons undergoes a mass shift of about $-200$ and $-250$ MeV, respectively. However, there was no reference for the density
 of the medium at which above mass shift will take place. From our present work within chiral SU(3) model and QCD sum rules and its comparison with earlier work shows that the drop in the masses of $D$ mesons at higher values of baryonic density  is sensitive to the model calculations. Since the medium modifications of decay width of charmonium states are found to be very sensitive to the medium modification of $D$ mesons masses and therefore, it is important to explore experimentally the densities at which above much mass shift of $D$ mesons will occur. The CBM experiment of FAIR project, where hadronic matter at high baryonic densities may be produced, may remove the above uncertainties in the calculations within different models.
 
In addition to the decay of charmonium to $D\bar{D}$ pairs, there can be contribution to decay width through the scattering of charmonium by nucleons e.g., $\psi$ + $N$ $\rightarrow$ $\Lambda_c$ + $\bar{D}$.  For $\psi(3686)$ decay width is written as \cite{friman}
\begin{eqnarray}
\Gamma_N=\frac{1}{\tau} =\langle \sigma_{abs}(\psi(3686)+N)
v_{rel} \rho_B \rangle,
\label{eq_decay_psi_n}
\end{eqnarray}
where $\sigma_{abc}$($\psi$3686 + N) is the cross section of $\psi$(3686) state with nucleon and has value  $\approx$ $6mb$ and $v_{rel}$ represents the average relative velocity  of nucleons in the initial state. In the rest frame of medium $v_{rela}$ is given by
 $v_{rel}=\frac{3}{4}
\frac{p_F}{m_N^{\star}}[1+\frac{2}{3}(\frac{m_N^{\star} p_\psi}{m_\psi p_F})^2]$,
where $p_F$ is the Fermi momentum of nuclear matter, $p_{\psi}$ is the momentum of charmonium and $m_N^{\star}$ is the in-medium mass of nucleons.
Allowing the medium modification of nucleon mass within chiral SU(3) model through expression $m_{N}^{\star} = -g_{\sigma N} \sigma  - g_{\zeta N} \zeta$, where $g_{\sigma N}$ and $g_{\sigma N}$ are coupling constants having value $10.6$ and $-0.47$, respectively, the decay width of $\psi(3686)$, using $p_{\psi} = 0$, at $\rho_B = \rho_0$ (4$\rho_0$), are observed to be 5.8 (79) and 5.6 (78) MeV at $f_s = 0$  and 0.5, respectively. However, if one does not consider the medium modification of nucleons, then the 
values of decay width at $\rho_B = \rho_0$ (4$\rho_0$) will be 3.6 (23) MeV, for $f_s = 0$.

In future, it will be interesting to investigate the effects of medium modification of $B$ mesons calculated in the present work on the in-medium decay width of bottomonium states. Also, the medium modifications of pseudoscalar charm strange and bottom strange mesons and their in-medium decay width will be investigated in near future.

\section{Summary}
\label{sec_summary}
In short, in the present investigation, we studied the in medium modification of the masses and decay constants of the pseudoscalar $D$ and $B$ mesons in  isospin asymmetric strange hadronic medium at finite temperature using the chiral SU(3) model and QCD sum rule approach. We observe a negative shift in the values   of the masses and decay constants for $D$ and $B$ mesons 
as a function of density of the medium.  The isospin asymmetry of the medium causes the mass splitting in isospin $D(D^+, D^0)$ and $B(B^+, B^0)$ meson doublets. The finite strangeness fraction causes decrease, whereas the finite temperature causes an increase in the values of the masses and decay constants of $D$ and $B$ mesons. The observed effect of the isospin asymmetry of the medium on the mass modifications of $D$ and $B$  mesons can be verified experimentally through the ratio $\frac{D^+}{D^0}$ and  $\frac{B^+}{B^0}$ in heavy-ion collision experiments.
The effects of medium modification of  $D$ mesons on the strong decay width of $\psi(3686)$, $\psi(3770)$ and $\chi(3556)$,  states  were calculated using the $^3 P_0$ model.
The in-medium decay width of charmonium states are found to be very sensitive to the in-medium masses of $D$ mesons and this suggest the importance of considering the internal structure of $D$ mesons for calculating the decay width. These observations may have important implications on
$J/\psi$ suppression in heavy-ion collision experiments.  
 These medium modification of the masses, decay constants of $D$ and $B$ mesons and decay width of the higher charmonium states can be experimentally verified in the CBM and PANDA experiments of  FAIR project at GSI, Germany.

\acknowledgements
 The authors gratefully acknowledge the financial support from
the Department of Science and Technology (DST), Government of India for research project under
 fast track scheme for young scientists (SR/FTP/PS-209/2012).

\end{document}